\documentclass[12pt,a4paper]{article}

\usepackage{jstyle}
\usepackage{natbib}
\usepackage[utf8]{inputenc}
\usepackage[english]{babel}

\usepackage{hyperref}
\usepackage{amsmath,amsfonts,amssymb,graphicx,color,mathtools}
\usepackage{axodraw}

\newcommand{\id}{\text{id}} \newcommand{\I}{\hat{\mathbb{I}}} \newcommand{\sgn}{\text{sgn}}
\newcommand{\res}{\text{res}} \newcommand{\Hom}{\text{Hom}} \newcommand{\amp}{\text{amp}}

\title{Log-Expansions from Combinatorial Dyson-Schwinger Equations} 
\author{Olaf Kr\"uger}
\affiliation{University of Vienna, Faculty of Physics, Boltzmanngasse 5, 1090 Vienna, Austria}
\emailAdd{olaf.krueger@univie.ac.at}

\abstract{We give a precise connection between combinatorial Dyson-Schwinger equations and log-expansions for
  Green's functions in quantum field theory. The latter are triangular power series in the coupling constant
  $\alpha$ and a logarithmic energy scale $L$ --- a reordering of terms as
  $G(\alpha,L) = 1 \pm \sum_{j \geq 0} \alpha^j H_j(\alpha L)$ is the corresponding log-expansion. In a first
  part of this paper, we derive the leading-log order $H_0$ and the next-to$^{(j)}$-leading log orders $H_j$
  from the Callan-Symanzik equation. In particular, $H_j$ only depends on the $(j+1)$-loop $\beta$-function and
  anomalous dimensions. For the photon propagator Green's function in quantum electrodynamics (and in a toy
  model, where all Feynman graphs with vertex sub-divergences are neglected), our formulas reproduce the known
  expressions for the next-to-next-to-leading log approximation in the literature. In a second part of this
  work, we review the connection between the Callan-Symanzik equation and Dyson-Schwinger equations, i.e.
  fixed-point relations for the Green's functions. Combining the arguments, our work provides a derivation of
  the log-expansions for Green's functions from the corresponding Dyson-Schwinger equations.}

\begin{document}

\begin{flushright}
  UWThPh 2019-18
\end{flushright}

\maketitle

\section{Prologue: Green's Functions in Quantum Field Theory}
\label{sec:introduction}

Quantum field theory (QFT) predicts probabilities for certain particle processes. For example, initialize an
experiment, where a virtual photon decays into an electron-positron pair. One cannot predict if and when the
photon decays, and if it does, where the electron and positron will go in the end. However, one can put two
detectors $D_1$ and $D_2$ somewhere and predict the probability, that an electron enters $D_1$ and a positron
enters $D_2$. This situation is totally different from deterministic classical mechanics, which predicts the
exact time evolution for given initial conditions.

Such probabilities can be found in the following way: Each QFT is defined by a Lagrangian, which is a functional
of the described particle fields. It consists of a kinetic part (terms that are quadratic in the fields) and
interaction terms (of cubic or higher order in the fields) --- from all these, one can read off Feynman rules in
a simple way. Now, a given particle process corresponds to a diagram (called Feynman graph) that translates via
these rules to the so-called Feynman amplitude. Finally, the latter translates via the LSZ formula to a
probability amplitude that squares to the actual probability~\footnote{This is similar to electrodynamics, where
  the $E$ field is an amplitude and only the intensity $I \propto E^2$ can be measured.}.

Consider for example the experiment above. The QFT that describes this process is quantum electrodynamics (QED).
The corresponding Lagrangian is a functional of the photon field $A^\mu$, the electron $\psi$ and its
anti-particle $\bar \psi$ (i.e. the positron). It contains one kinetic term for the photon and one for the
electron and positron together; as well as one interaction term, i.e.
$g\bar{\psi} A^\mu \gamma_\mu \psi$~\footnote{As a convenient toy model, we consider QED in \textit{Landau
    gauge}. Therefore, another term in the Lagrangian vanishes \cite{Kissler:2018lnn} and the theory contains
  only one coupling constant. We discuss more general cases in Section~\ref{sec:gener-results}.}. Here, the
\textit{coupling constant} $g \ll 1$ is the electric charge of the electron and $\gamma_\mu$ is a Dirac matrix.
Each process involving these particles translates to a graph that consists of edges and vertices: A photon
corresponds to a wiggly line and a straight arrow-line indicates an electron or positron (depending on the
direction of the arrow). An interaction between these three particles is represented by a vertex.
Table~\ref{tab:Residues} shows all these basic ingredients of QED Feynman graphs. The virtual photon decay
described above corresponds to the vertex graph in Table~\ref{tab:Residues} and Feynman rules state that the
respective Feynman amplitude is proportional to the electric charge $g$. It also depends on other factors, e.g.
on the relative positions of the detectors.

Quantum mechanics now tells us that the obtained result is \textit{not exact}. Indeed, one has to consider all
possible ways in which the final state (an electron in $D_1$ and a positron in $D_2$) is achieved. For example,
both particles could actually interchange another photon before entering the detectors. The respective Feynman
diagram is the 1-loop graph
\begin{equation}
\label{eq:ex1}
  \SetScale{0.5}
  \begin{picture}(40,24)(0,16)
    \Photon(0,36)(32,36){3.3}{3.5} \Vertex(32,36){1.7}
    \Line(80,72)(32,36) \ArrowLine(64,12)(32,36)
    \Line(80,0)(32,36) \ArrowLine(32,36)(64,60)
    \Photon(64,12)(64,60){-3.3}{4.5} \Vertex(64,60){1.7}\Vertex(64,12){1.7}
\end{picture}\,, \vspace{0.4cm}
\end{equation}
and one has to add the resulting Feynman amplitude to the previous one before computing the actual probability.
Note that the latter amplitude is of order $g^3$, because by Feynman rules, each vertex constitutes one factor
of $g$. Thus, the contribution to the overall probability is small compared to the initial one. All in all, one
has to consider infinitely many Feynman graphs with more and more loops and vertices, whose contributions to the
final amplitude (the \textit{quantum corrections}) become smaller and smaller. At some point, one may truncate
the resulting Feynman amplitude when it is accurate enough.

There are two problems in this calculation: First, each loop diagram translates via Feynman rules to a
\textit{divergent} integral. In order to extract the correct quantum corrections, one has to \textit{regularize}
the integrals, i.e. to keep track of the different kinds of divergences. Then, one applies a
\textit{renormalization scheme}, i.e. one introduces `Counter terms' into the Lagrangian, which lead to
additional Feynman graphs, whose amplitudes exactly cancel the divergences. In this way, the resulting Feynman
amplitudes become finite~\footnote{Another way to think about regularization and renormalization is the
  following: The infinities are absorbed by the fields and coupling constants in the Lagrangian. In this way,
  they become unphysical, but the resulting probability amplitudes remain finite and are physically
  observable.}. A second problem are \textit{infrared divergences} that occur when the probability of a particle
process is computed from the Feynman amplitude. We do not discuss these any further, because our work is focused
on the computation of Feynman amplitudes.

\begin{table}[t]
  \centering
  \begin{tabular}{ccc}
    Electron and Positron & $\qquad$ Photon$\qquad $ & Interaction vertex\\
    \SetScale{0.5}
    \begin{picture}(30,36)(0,8) \ArrowLine(0,36)(60,36)\end{picture} &
                                                                       \SetScale{0.5} \begin{picture}(30,36)(0,8)                                                       \Photon(0,36)(60,36){3.3}{6.5}\end{picture}
                                                     &  \SetScale{0.5} \begin{picture}(30,36)(0,8)
                                                       \Photon(0,36)(32,36){3.3}{3.5} \Vertex(32,36){1.7}
                                                       \ArrowLine(32,36)(60,64) \ArrowLine(60,8)(32,36)
                                                     \end{picture}   
  \end{tabular}
  \caption{The different types of edges and vertices in QED Feynman graphs.}
  \label{tab:Residues}
\end{table}

After regularization and renormalization in a momentum subtraction scheme, each Feynman graph $\Gamma$
contributes a quantum correction $\phi_R(\Gamma)$ to the initial Feynman amplitude. This correction depends on
the coupling constants $g_k$ and the scalar products $p_i \cdot p_j$ between the momenta $p^\mu_i$ of the
particles that enter the process. Here, it is very convenient to factor out an energy scale $S$ and to define
dimensionless \textit{scattering angles} as $\mathbf{\Theta} = \{p_i \cdot p_j \slash S\}$. Now, renormalization
requires some sort of boundary condition. In our case, we assume that $\phi_R(\Gamma)$ is known for certain
values of $S = S_0$ and $\mathbf{\Theta} = \mathbf{\Theta}_0$. One usually calls $S_0$ \textit{renormalization
  scale} and $\{S_0,\mathbf{\Theta}_0\}$ \textit{renormalization point}~\footnote{In quantum chromodynamics
  (QCD), there are other kinds of boundary conditions. However, this does not alter the structure of
  $\phi_R(\Gamma)$ that is discussed in the following.}. Then, it turns out that $\phi_R(\Gamma)$ is a
polynomial in the coupling constants $g_k$ and the logarithm $L = \log (S \slash S_0)$ of the energy scale
\cite{Brown:2011pj}. Hence, \textit{renormalized Feynman rules} can be written as a linear map $\phi_R$ from the
set of Feynman graphs to the polynomial algebra $\mathcal{A}[g_k,L]$, where the dependence of $\phi_R(\Gamma)$
on the scattering angles $\mathbf{\Theta}$ and $\mathbf{\Theta}_0$ is hidden in the coefficients of the
polynomial. In the following, we will make this more explicit.

First, $\phi_R$ is an \textit{algebra homomorphism} --- the domain of definition is the algebra $\mathcal{H}$ of
all \textit{one-particle irreducible (1PI)} Feynman loop graphs~\footnote{1PI Feynman graphs remain connected
  when one internal edge is removed. For example,
  $\SetScale{0.5}\begin{picture}(40,18)(0,15) \Photon(0,36)(20,36){3.3}{2.5} \Vertex(20,36){1.7}
    \Photon(60,36)(80,36){3.3}{2.5} \Vertex(60,36){1.7} \LongArrowArcn(40,36)(20,35,125)
    \LongArrowArcn(40,36)(20,125,215) \LongArrowArcn(40,36)(20,215,305) \LongArrowArcn(40,36)(20,305,395)
    \Vertex(40,16){1.7} \Vertex(40,56){1.7} \Photon(40,16)(40,56){3.3}{4}
  \end{picture}$ is 1PI, while
  $\SetScale{0.5}\begin{picture}(70,18)(0,15) \Photon(0,36)(20,36){3.3}{2.5} \Photon(60,36)(80,36){3.3}{2.5}
    \Photon(120,36)(140,36){3.3}{2.5} \Vertex(20,36){1.7} \Vertex(60,36){1.7} \LongArrowArcn(40,36)(20,80,260)
    \LongArrowArcn(40,36)(20,260,440) \Vertex(80,36){1.7} \Vertex(120,36){1.7} \LongArrowArcn(100,36)(20,35,125)
    \LongArrowArcn(100,36)(20,125,215) \LongArrowArcn(100,36)(20,215,305) \LongArrowArcn(100,36)(20,305,395)
    \Vertex(100,16){1.7} \Vertex(100,56){1.7} \Photon(100,16)(100,56){3.3}{4}
  \end{picture}$
  is not. The latter is called one-particle reducible (1PR).\\[-5pt]} (of a given QFT) and their disjoint unions
(The disjoint union is an associative product and the empty graph $\mathbb{I}$ is the corresponding unit
element). Furthermore, Feynman rules are normalized such that a 1PI Feynman graph $\Gamma$ contributes a
change of $\phi_R(\Gamma)$ relative to the corresponding zero-loop Feynman amplitude~\footnote{In
  this way, $\phi_R(\Gamma)$ is a scalar, although the corresponding amplitude may be of a tensorial structure
  (e.g. it could be proportional to a Dirac matrix $\gamma_\mu$).}. This implies that 1PI graphs without loops
would map to $1$ (the unit element in $\mathcal{A}$). Note that these are not in $\mathcal{H}$ but are
identified with the empty graph $\mathbb{I}$. For example, the QED graphs in Table~\ref{tab:Residues} would map
to $1$ by renormalized Feynman rules~\footnote{There is no problem, that three different graphs are identified 
  with the same element $\mathbb{I}$. All of them are mapped to $1$, so why distinguish?} and the 1-loop graph
in Eq.~(\ref{eq:ex1}) maps to a term that is proportional to $g^2$ (and not $\propto g^3$). With this
normalization, a one-particle reducible graph (1PR) corresponds to the disjoint union of its 1PI parts, e.g.
\begin{equation}
\label{eq:ex2}
\SetScale{0.5}
\phi_R 
\Big(
\begin{picture}(70,24)(0,15) \Photon(0,36)(20,36){3.3}{2.5} \Photon(60,36)(80,36){3.3}{2.5}
\Photon(120,36)(140,36){3.3}{2.5} \Vertex(20,36){1.7} \Vertex(60,36){1.7}
\LongArrowArcn(40,36)(20,80,260) \LongArrowArcn(40,36)(20,260,440)
\Vertex(80,36){1.7} \Vertex(120,36){1.7} \LongArrowArcn(100,36)(20,35,125) 
\LongArrowArcn(100,36)(20,125,215) \LongArrowArcn(100,36)(20,215,305) \LongArrowArcn(100,36)(20,305,395) 
\Vertex(100,16){1.7} \Vertex(100,56){1.7} \Photon(100,16)(100,56){3.3}{4}
\end{picture}
\Big)
=
\phi_R
\Big(
\begin{picture}(40,24)(0,15)
\Photon(0,36)(20,36){3.3}{2.5} \Vertex(20,36){1.7} 
\Photon(60,36)(80,36){3.3}{2.5} \Vertex(60,36){1.7}
\LongArrowArcn(40,36)(20,80,260) \LongArrowArcn(40,36)(20,260,440)
\end{picture}  \;
\begin{picture}(40,24)(0,15)
\Photon(0,36)(20,36){3.3}{2.5} \Vertex(20,36){1.7} 
\Photon(60,36)(80,36){3.3}{2.5} \Vertex(60,36){1.7}
\LongArrowArcn(40,36)(20,35,125) \LongArrowArcn(40,36)(20,125,215)
\LongArrowArcn(40,36)(20,215,305) \LongArrowArcn(40,36)(20,305,395)
\Vertex(40,16){1.7} \Vertex(40,56){1.7} \Photon(40,16)(40,56){3.3}{4}
\end{picture}
\Big)
=
\phi_R
\Big(
\begin{picture}(40,24)(0,15)
\Photon(0,36)(20,36){3.3}{2.5} \Vertex(20,36){1.7} 
\Photon(60,36)(80,36){3.3}{2.5} \Vertex(60,36){1.7}
\LongArrowArcn(40,36)(20,80,260) \LongArrowArcn(40,36)(20,260,440)
\end{picture}  
\Big)  
\phi_R
\Big(
\begin{picture}(40,24)(0,15)
\Photon(0,36)(20,36){3.3}{2.5} \Vertex(20,36){1.7} 
\Photon(60,36)(80,36){3.3}{2.5} \Vertex(60,36){1.7}
\LongArrowArcn(40,36)(20,35,125) \LongArrowArcn(40,36)(20,125,215)
\LongArrowArcn(40,36)(20,215,305) \LongArrowArcn(40,36)(20,305,395)
\Vertex(40,16){1.7} \Vertex(40,56){1.7} \Photon(40,16)(40,56){3.3}{4}
\end{picture}
\Big)\,. \vspace{0.3cm}
\end{equation}

Secondly, the algebra $\mathcal{H}$ acquires a grading. In general, it is graded by the number of vertices of
each kind in a graph. However, for QFT's with only one interaction term, it is more convenient to grade the
algebra $\mathcal{H}$ by the number of loops,
\begin{equation}
  \label{eq:grading}
  \mathcal{H} = \bigoplus_{k \geq 0} \mathcal{H}_k\,.
\end{equation}
For example, $\mathcal{H}_0 = \{\mathbb{I}\}$, the graph in Eq.~(\ref{eq:ex1}) belongs to $\mathcal{H}_1$ and
the disjoint union in Eq.~(\ref{eq:ex2}) is an element in $\mathcal{H}_3$ etc. In this case, there exists a
redefinition of the coupling constant as $\alpha = g^l$, such that renormalized Feynman rules map each $k$-loop
graph to a term proportional to $\alpha^k$. We show this explicitly in Section~\ref{sec:grading-mathcalh}. For a
vertex with three outgoing edges (e.g. QED), one finds $l = 2$, whereas for a vertex with four outgoing edges
(e.g. in $\phi^4$ theory), $l = 1$. For simplicity, we state our results using this redefinition. The
generalization to QFT's with more than one coupling constant is presented in Section~\ref{sec:gener-results}.

All in all, renormalized Feynman rules are an algebra homomorphism
\begin{equation*}
  \phi_R : \mathcal{H} \to \mathcal{A}[\alpha,L]\,,
\end{equation*}
and the quantum corrections of a certain particle process are found by applying $\phi_R$ to an infinite sum of
graphs in $\mathcal{H}$ (1PI graphs and disjoint unions thereof). This results in a series expansion in the
coupling constant $\alpha$ and the logarithm $L$, where the dependence on the scattering angles
$\mathbf{\Theta}$ and $\mathbf{\Theta_0}$ is hidden in the coefficients.

With this definition of renormalized Feynman rules, we will shortly explain the notion of Green's functions, for
which we need some more notations: The \textit{residue} $r$ of a Feynman graph is its external leg structure.
For example, the graph in Eq.~(\ref{eq:ex1}) has residue
$r = \SetScale{0.5} \begin{picture}(15,0)(0,6) \Photon(0,18)(16,18){2}{2.5} \Vertex(16,18){1.7}
  \ArrowLine(16,18)(30,32) \ArrowLine(30,4)(16,18) \end{picture}$
and those in Eq.~(\ref{eq:ex2}) have residue
$r = \SetScale{0.5} \begin{picture}(11,0)(0,6) \Photon(0,18)(20,18){2}{3.5}\end{picture}$. Furthermore, its
degree $|r|$ denotes the number of external edges --- Feynman diagrams with $|r| = 2$ are called `propagator
graphs' and those with $|r| > 2$ `vertex graphs'. Finally, we define $\sgn(r)$, which is $+1$ for vertex- and
$-1$ for propagator graphs.

Now, in order to compute the exact Feynman amplitude of a certain particle process, one has to find the
\textit{Green's functions} of the respective QFT. These encode all necessary information about the full quantum
corrections of all particle processes under consideration. For example, consider another QED experiment: A
source $S$ emits an electron, which is absorbed at a later time by a detector $D$. The simplest way for the
electron would be to go straight from $S$ to $D$. Then, the quantum corrections to the corresponding Feynman
amplitude are obtained by applying renormalized Feynman rules to the sum of all Feynman diagrams with residue
$r = \SetScale{0.5} \begin{picture}(11,0)(0,6) \ArrowLine(0,18)(20,18)\end{picture}$. We denote this sum by an
ellipse. It is given by
\begin{align}
\label{eq:geometric}
\SetScale{0.5}
\begin{picture}(40,0)(0,15)
\ArrowLine(0,36)(15,36) \ArrowLine(65,36)(80,36) \GOval(40,36)(15,25)(0){0.5}
\end{picture} 
=
\begin{picture}(30,0)(0,15)
\ArrowLine(0,36)(60,36)
\end{picture} +
\begin{picture}(40,0)(0,15)
\ArrowLine(0,36)(20,36) \ArrowLine(60,36)(80,36) \GCirc(40,36){20}{0.5}
\end{picture} +
\begin{picture}(70,0)(0,15)
\ArrowLine(0,36)(20,36) \ArrowLine(60,36)(80,36) \ArrowLine(120,36)(140,36)
\GCirc(40,36){20}{0.5} \GCirc(100,36){20}{0.5} 
\end{picture} + \ldots \,,
\end{align}
where the circle denotes the sum of all 1PI loop-graphs. Thus, thanks to the geometric series,
\begin{align*} \SetScale{0.5}
  \phi_R \Big(
  \begin{picture}(40,0)(0,15)
    \ArrowLine(0,36)(15,36) \ArrowLine(65,36)(80,36) \GOval(40,36)(15,25)(0){0.5}
  \end{picture} 
  \Big) = \Big[ 1 - \phi_R \Big( \begin{picture}(40,0)(0,15)
\ArrowLine(0,36)(20,36) \ArrowLine(60,36)(80,36) \GCirc(40,36){20}{0.5}
\end{picture} \Big) \Big]^{-1}\,.
\end{align*}
This works for all propagator graphs. Hence, let
\begin{equation}
  \label{eq:X-propagator}
  X^r = \mathbb{I} - \sum\text{1PI loop graphs with residue } r\,, \qquad |r| = 2 \,.
\end{equation}
Then, the corresponding \textit{propagator Green's function} $G^r$ is defined as $G^r(\alpha,L) = \phi_R(X^r)$ .
It is the inverse of the quantum corrections to the respective 1-particle process (the particle goes from $S$ to
$D$).

For interaction processes, the situation is similar, but there is no geometric series to compute. Therefore, one
simply defines
\begin{equation}
  \label{eq:X-vertex}
  X^r = \mathbb{I} + \sum\text{1PI loop graphs with residue } r\,, \qquad |r| > 2 \,,
\end{equation}
and calls $G^r$ with $G^r(\alpha,L) = \phi_R(X^r)$ the \textit{vertex Green's function of residue $r$}. Knowing
these functions allows to compute the correct Feynman amplitude of any interaction process that is described by
the theory. Let us illustrate this again at the example of QED. First, consider the photon decay described at
the beginning. The full quantum corrections are given by the Green's functions as follows:
\begin{equation}
\label{eq:ex3}
    \SetScale{0.5}
    \phi_R \left(
    \begin{picture}(72,35)(0,26)
      \Photon(0,57)(16,57){3.3}{1.5} \GOval(32,57)(10,16)(0){0.5} \Photon(48,57)(64,57){3.3}{1.5}
      \GCirc(80,57){16}{0.5} 
      \ArrowLine(91.3,68.3)(102.6,79.6) \GOval(113.9,90.9)(10,16)(45){0.5} \ArrowLine(125.2,102.2)(136.5,113.5)
      \ArrowLine(102.6,34.4)(91.3,45.7) \GOval(113.9,23.1)(10,16)(-45){0.5} \ArrowLine(136.5,0.5)(125.2,11.8)
    \end{picture} \right) = \phi_R
  \left(
    \begin{picture}(30,14)(0,26)
      \Photon(0,57)(16,57){3.3}{1.5} \GCirc(32,57){16}{0.5} 
      \ArrowLine(43.3,68.3)(54.6,79.6) \ArrowLine(54.6,34.4)(43.3,45.7)
    \end{picture}
  \right)\phi_R 
  \left(
    \begin{picture}(32,0)(0,26)
      \Photon(0,57)(16,57){3.3}{1.5} \GOval(32,57)(10,16)(0){0.5} \Photon(48,57)(64,57){3.3}{1.5}
    \end{picture}
  \right) \phi_R \left(
    \begin{picture}(32,0)(0,26)
      \ArrowLine(0,57)(16,57) \GOval(32,57)(10,16)(0){0.5} \ArrowLine(48,57)(64,57)
    \end{picture} \right)^2  = 
  \SetScale{0.3} \frac{G^{\begin{picture}(9,0)(0,0) \Photon(0,18)(16,18){2}{2.5} \Vertex(16,18){1.7}
        \ArrowLine(16,18)(30,32) \ArrowLine(30,4)(16,18) \end{picture}}(\alpha,L)}{G^{\begin{picture}(6,0)(0,0)
        \Photon(0,18)(20,18){2}{3.5}\end{picture}}(\alpha,L) G^{\begin{picture}(6,0)(0,0)
        \ArrowLine(0,18)(20,18)\end{picture}}(\alpha,L)^2} \,. 
\end{equation}
Here, the circle on the vertex denotes the sum of all 1PI graphs \textit{including} the zero-loop graph, in
contrast to the `propagator circles'. Indeed, the argument of $\phi_R$ on the lhs above denotes the sum of all
Feynman graphs with residue
$r = \SetScale{0.5} \begin{picture}(15,0)(0,6) \Photon(0,18)(16,18){2}{2.5} \Vertex(16,18){1.7}
  \ArrowLine(16,18)(30,32) \ArrowLine(30,4)(16,18) \end{picture}$.
Secondly, consider a scattering between two electrons. The corresponding Feynman graphs have four external legs
and in this case, the zero-loop graph
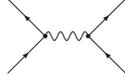
\begin{equation*}
  \SetScale{0.5} \begin{picture}(44,12)(0,15) \Photon(28,36)(60,36){3.3}{3.5} \Vertex(28,36){1.7}
    \ArrowLine(28,36)(0,64) \ArrowLine(0,8)(28,36) \Vertex(60,36){1.7} \ArrowLine(60,36)(88,64)
    \ArrowLine(88,8)(60,36) \end{picture}
\end{equation*}
is not 1PI. But again, the full quantum corrections to the corresponding Feynman amplitude are given by the
Green's functions:
\begin{equation}
\label{eq:ex4}
  \SetScale{0.5}
  \phi_R \left(\,
    \begin{picture}(60,35)(0,26)
      \GCirc(56.5,57){16}{0.5} 
      \ArrowLine(45.2,68.3)(33.9,79.6) \GOval(22.6,90.9)(10,16)(-45){0.5} \ArrowLine(11.3,102.2)(0,113.5)
      \ArrowLine(33.9,34.4)(45.2,45.7) \GOval(22.6,23.1)(10,16)(45){0.5} \ArrowLine(0,0.5)(11.3,11.8)
      \ArrowLine(67.8,68.3)(79.1,79.6) \GOval(90.4,90.9)(10,16)(45){0.5} \ArrowLine(101.7,102.2)(113,113.5)
      \ArrowLine(79.1,34.4)(67.8,45.7) \GOval(90.4,23.1)(10,16)(-45){0.5} \ArrowLine(113,0.5)(101.7,11.8)
    \end{picture} + \,
    \begin{picture}(108,35)(0,26)
      \Photon(72.5,57)(88.5,57){3.3}{1.5} \GOval(104.5,57)(10,16)(0){0.5} \Photon(120.5,57)(136.5,57){3.3}{1.5}
      \GCirc(152.5,57){16}{0.5} \GCirc(56.5,57){16}{0.5} 
      \ArrowLine(45.2,68.3)(33.9,79.6) \GOval(22.6,90.9)(10,16)(-45){0.5} \ArrowLine(11.3,102.2)(0,113.5)
      \ArrowLine(33.9,34.4)(45.2,45.7) \GOval(22.6,23.1)(10,16)(45){0.5} \ArrowLine(0,0.5)(11.3,11.8)
      \ArrowLine(163.8,68.3)(175.1,79.6) \GOval(186.4,90.9)(10,16)(45){0.5} \ArrowLine(197.7,102.2)(209,113.5)
      \ArrowLine(175.1,34.4)(163.8,45.7) \GOval(186.4,23.1)(10,16)(-45){0.5} \ArrowLine(209,0.5)(197.7,11.8)
    \end{picture}
  \right) = \SetScale{0.3} \frac{G^{\begin{picture}(9,0)(0,0) \ArrowLine(16,18)(2,32) \ArrowLine(2,4)(16,18)
        \Vertex(16,18){1.7} \ArrowLine(16,18)(30,32) \ArrowLine(30,4)(16,18) \end{picture}}(\alpha,L) +
    G^{\begin{picture}(9,0)(0,0) \Photon(0,18)(16,18){2}{2.5} \Vertex(16,18){1.7} \ArrowLine(16,18)(30,32)
            \ArrowLine(30,4)(16,18) \end{picture}}(\alpha,L)^2 \big\slash G^{\begin{picture}(6,0)(0,0)
          \Photon(0,18)(20,18){2}{3.5}\end{picture}}(\alpha,L)}{G^{\begin{picture}(6,0)(0,0)
        \ArrowLine(0,18)(20,18)\end{picture}}(\alpha,L)^4} \,.
\end{equation}
Note that the argument of $\phi_R$ is the sum of all Feynman graphs with residue
$\SetScale{0.5} \begin{picture}(15,0)(0,6) \ArrowLine(16,18)(2,32) \ArrowLine(2,4)(16,18) \Vertex(16,18){1.7}
  \ArrowLine(16,18)(30,32) \ArrowLine(30,4)(16,18) \end{picture}$.
As before, the circle on the vertex with four outgoing edges denotes the sum of all corresponding 1PI graphs.
However, it does \textit{not} contain the respective zero-loop Feynman graph, because it does not
exist~\footnote{A comment on the notation: Circles denote the sum of all 1PI graphs with the respective residue.
  For propagators, the zero-loop graph is \textit{excluded}, while for vertices, it is \textit{included} (if it
  exists).}.

Let us summarize: One can compute the Feynman amplitude of any particle process when the Green's functions of
the corresponding QFT are known. These are a power series in the coupling constant $\alpha$ and the logarithm of
the energy scale $L = \log S \slash S_0$. Hence, $G^r \in \mathcal{A}[\alpha,L]$. But renormalized Feynman rules
tell us slightly more, namely that a $k$-loop graph maps to a polynomial of degree smaller or equal than $k$ in
$L$. Therefore, $G^r$ can be written as a triangular power series,
\begin{equation}
  \label{eq:G-perturb}
  G^r(\alpha,L) = \phi_R(r) + \sgn(r) \sum_{k\geq 1} \alpha^k G^r_k(L)\,,
\end{equation}
where $G^r_k$ is a degree-$k$ polynomial that encodes the contributions of all $k$-loop graphs to
$G^r$~\footnote{The $\sgn(r)$ is a convention that is related to the different signs in the definitions of $X^r$
  in Eqs.~(\ref{eq:X-propagator}) and (\ref{eq:X-vertex}).}. In particular, these polynomials vanish at the
renormalization scale (for $L = \log S \slash S_0 =  0$), such that
\begin{equation*}
  G^r(\alpha,0) = \phi_R(r) =
  \begin{cases}
    1, & \text{if the 1PI zero-loop graph $r$ exists,}\\ 0,& \text{else}
  \end{cases}.
\end{equation*}

The Green's functions are infinite power series and the best one can do is to approximate them perturbatively.
One possibility is to truncate the sum in Eq.~(\ref{eq:G-perturb}) at some loop-order. Since $\alpha \ll 1$,
this is a reasonable approach, at least for energies near the renormalization scale $S \sim S_0$. However, the
perturbative computation breaks down at energies far away from $S_0$, where $L \gg 1$. Then, it is better to
reorder the terms in Eq.~(\ref{eq:G-perturb}) as follows:
\begin{equation}
\label{eq:H-log-exp}
  G^r(\alpha,L) = H^r(\alpha,\alpha L), \qquad H^r(\alpha,z) = \phi_R(r) + \sgn(r) \sum_{j \geq 0} \alpha^j
  H^r_j(z)\,. 
\end{equation}
This is called the \textit{log-expansion} for the Green's function $G^r$. $H^r_0$ is the \textit{leading log
  order} (LLO), $H^r_1$ is the \textit{next-to-leading log order} (NLLO) and in general, $H^r_j$ is the
\textit{next-to$^{(j)}$-leading log order} (N$^{(j)}$LLO). A truncation of the above sum, such that terms of
order $\mathcal{O}(\alpha^{n+1})$ are neglected is called the \textit{next-to$^{(n)}$-leading log (N$^{(n)}$LL)
  approximation} of $H^r$. Perturbatively, this gives accurate results, as long as the functions $H^r_j$ are
regular at $z = \alpha L$, which may well be the case for energies even far away from the renormalization scale.

\textbf{In this work, we present the following results:}
\begin{itemize}
\item First, we derive the N$^{(j)}$LL approximation of the Green's functions $H^r$ in Eq.~(\ref{eq:H-log-exp})
  with $\phi_R(r) = 1$~\footnote{These are the propagator Green's functions and the vertex Green's function that
    belongs to the interaction term in the Lagrangian.} from the \textit{Callan-Symanzik equation}. The latter
  is a simple first-order partial differential equation for $G^r$ that describes its dependence on the energy
  scale $L$, see Eq.~(\ref{eq:MAIN}). In particular, $H^r_j$ only depends on the $(j+1)$-loop $\beta$-function
  and anomalous dimension, which are defined in Section~\ref{sec:anom-dimens-beta} and only depend on the
  Feynman amplitudes of at most $(j+1)$-loop graphs. In principle, this calculation must be long known in
  particle physics. However, this is barely exhibited in sufficient clarity.
\item As an application, we compute the NNLL approximation for the propagator Green's functions in a class of
  QFT toy models, in which Feynman graphs with vertex sub-divergences are neglected. An example is the photon
  propagator Green's function
  $\SetScale{0.3} H^{\begin{picture}(6,0)(0,0) \Photon(0,18)(20,18){2}{3.5}\end{picture}}$ in QED, thanks to the
  Ward-Takahashi identities (see e.g. \cite{Ward:1950xp,Green:1953te,Takahashi:1957xn}). For these toy models,
  NNLL approximations have already been derived using chord diagrams \cite{Yeats} or the shuffle Hopf algebra of
  words \cite{Krueger:2014poa}. Our results (see Eq.~(\ref{eq:H_s})) agree with \cite{Yeats}, but differ from
  \cite{Krueger:2014poa}. The reason for the mismatch is a mistake in \cite{Krueger:2014poa}, where the shuffle
  products of commutators were computed using a wrong algorithm.
\item In the second part of this paper, we review the connection between the Callan-Symanzik equation and
  \textit{Dyson-Schwinger equations} (DSE's) \cite{Kreimer:2006ua,Kreimer:2006gm}, i.e. fixed-point relations
  for the infinite sums of Feynman graphs in Eqs.~(\ref{eq:X-propagator}) and (\ref{eq:X-vertex}). This can be
  seen by extending $\mathcal{H}$ to the \textit{Hopf algebra of Feynman graphs}
  \cite{Kreimer:1997dp,Connes:1999yr} and writing the DSE's using so-called \textit{insertion operators}
  \cite{Kreimer:2005rw}, see Eq.~(\ref{eq:DSE})~\footnote{The leading-log approximation of the Green's functions
    has been already derived using DSE's and the Hopf-algebra of words \cite{Delage:2016aho}. Our results agree,
    see the first term in Eq.~(\ref{eq:NNLL-H}).}.
\item Finally, we comment on possible extensions of our results. For the Green's functions $H^r$ in
  Eq.~(\ref{eq:H-log-exp}) with $\phi_R(r) = 0$ (such as
  $\SetScale{0.3}H^{\begin{picture}(9,0)(0,0) \ArrowLine(16,18)(2,32) \ArrowLine(2,4)(16,18) \Vertex(16,18){1.7}
      \ArrowLine(16,18)(30,32) \ArrowLine(30,4)(16,18) \end{picture}}$
  in QED), DSE's are barely discussed in details in the literature. When these are found, our results can be
  generalized in a simple way to find the corresponding log-expansions. On the other hand, there are DSE's for
  Green's functions in QFT's with more than one interaction term
  \cite{Connes:1999yr,vanSuijlekom:2009zz,Kreimer:2012jw,Foissy:2015wqu} --- we generalize our results to such
  cases in Section~\ref{sec:gener-results}.
\end{itemize}

The paper is organized as follows: In the next Section, we derive the log-expansions of the Green's functions
$H^r$ with $\phi_R(r) = 1$ for QFT's with only one interaction term in the Lagrangian. We also give the example
of the propagator Green's functions in toy models that neglect Feynman graphs with vertex sub-divergences. The
results are obtained from the Callan-Symanzik equation~(\ref{eq:MAIN}). In Section~\ref{sec:review}, we rederive
that relation from Dyson-Schwinger equations for the infinite sums of Feynman graphs in
Eqs.~(\ref{eq:X-propagator}) and (\ref{eq:X-vertex}). There, we also review some known aspects about the
combinatorics of Feynman graphs (we extend $\mathcal{H}$ to a Hopf algebra and rederive the essential properties
of renormalized Feynman rules $\phi_R$). In Section~\ref{sec:gener-results}, we generalize our results to QFT's
with more than one interaction term in the Lagrangian and conclude in Section~\ref{sec:conclusions}.

\section{Results}
\label{sec:results}

In this section, we present the derivation of the log expansion $H^r$ in Eq.~(\ref{eq:H-log-exp}) with
$\phi_R(r) = 1$ from the Callan-Symanzik equation~(\ref{eq:MAIN}), see Section~\ref{sec:log-expansion}.
Therefore, we first introduce the anomalous dimension and the $\beta$-function in
Section~\ref{sec:anom-dimens-beta}, as they appear in the final results. Finally,
Section~\ref{sec:exampl-qed-trunc} gives a simple example of the log-expansion for propagator Green's functions
in QFT toy models, where all Feynman graphs with vertex sub-divergences are neglected.

\subsection{The Anomalous Dimensions and $\beta$-Function}
\label{sec:anom-dimens-beta}

For each vertex residue $r = v$, define
\begin{equation}
  \label{eq:amp-greens}
  X^v_\amp = \frac{X^v}{\prod_{e \sim v} \sqrt{X^e}}\,.
\end{equation}
Here, $e\sim v$ means that the edge $e$ is incident to the vertex $v$. The square root must be expanded in a
Taylor series, similar to the geometric series of propagator Feynman graphs. For example, in QED:
\begin{equation*}
  \SetScale{0.3}X_\amp^{\begin{picture}(9,0)(0,0) \Photon(0,18)(16,18){2}{2.5} \Vertex(16,18){1.7}
      \ArrowLine(16,18)(30,32) \ArrowLine(30,4)(16,18) \end{picture}} = \frac{X^{\begin{picture}(9,0)(0,0)
        \Photon(0,18)(16,18){2}{2.5} \Vertex(16,18){1.7} \ArrowLine(16,18)(30,32)
        \ArrowLine(30,4)(16,18) \end{picture}}}{\sqrt{X^{\begin{picture}(6,0)(0,0)
          \Photon(0,18)(20,18){2}{3.5}\end{picture}}} X^{\begin{picture}(6,0)(0,0)
        \ArrowLine(0,18)(20,18)\end{picture}}}= \,
  \SetScale{0.5} \hspace{-13pt}
  \begin{picture}(72,20)(0,26) \GOval(32,57)(10,16)(0){0.5}
    \Photon(48,57)(64,57){3.3}{1.5} \GCirc(80,57){16}{0.5} \ArrowLine(91.3,68.3)(102.6,79.6)
    \GOval(113.9,90.9)(10,16)(45){0.5} \ArrowLine(125.2,102.2)(136.5,113.5) \ArrowLine(102.6,34.4)(91.3,45.7)
    \GOval(113.9,23.1)(10,16)(-45){0.5} \ArrowLine(136.5,0.5)(125.2,11.8)
    \CBox(15,32)(32,67){White}{White} \CTri(93.9,3.1)(133.9,43.1)(136.5,0.5){White}{White}
    \CTri(133.9,70.9)(93.9,110.9)(136.5,113.5){White}{White} 
  \end{picture}\,,\qquad \SetScale{0.3}
  X_\amp^{\begin{picture}(9,0)(0,0) \ArrowLine(16,18)(2,32) \ArrowLine(2,4)(16,18) \Vertex(16,18){1.7}
      \ArrowLine(16,18)(30,32) \ArrowLine(30,4)(16,18) \end{picture}} = \frac{X^{\begin{picture}(9,0)(0,0)
        \ArrowLine(16,18)(2,32) \ArrowLine(2,4)(16,18) \Vertex(16,18){1.7}
        \ArrowLine(16,18)(30,32) \ArrowLine(30,4)(16,18) \end{picture}}}{\left(X^{\begin{picture}(6,0)(0,0)
          \ArrowLine(0,18)(20,18)\end{picture}}\right)^2} = \hspace{-5pt} \SetScale{0.5}
  \begin{picture}(60,20)(0,26)
    \GCirc(56.5,57){16}{0.5} 
    \ArrowLine(45.2,68.3)(33.9,79.6) \GOval(22.6,90.9)(10,16)(-45){0.5} \ArrowLine(11.3,102.2)(0,113.5)
    \ArrowLine(33.9,34.4)(45.2,45.7) \GOval(22.6,23.1)(10,16)(45){0.5} \ArrowLine(0,0.5)(11.3,11.8)
    \ArrowLine(67.8,68.3)(79.1,79.6) \GOval(90.4,90.9)(10,16)(45){0.5} \ArrowLine(101.7,102.2)(113,113.5)
    \ArrowLine(79.1,34.4)(67.8,45.7) \GOval(90.4,23.1)(10,16)(-45){0.5} \ArrowLine(113,0.5)(101.7,11.8)
    \CTri(70.4,3.1)(110.4,43.1)(113,0.5){White}{White} \CTri(110.4,70.9)(70.4,110.9)(113,113.5){White}{White} 
    \CTri(42.6,3.1)(2.6,43.1)(0,0.5){White}{White}
    \CTri(42.6,110.9)(2.6,70.9)(0,113.5){White}{White}
  \end{picture}\,. 
\end{equation*}
The square roots are graphically represented by half-ellipses. Since renormalized Feynman rules $\phi_R$ are an
algebra homomorphism, one can easily evaluate these infinite sums of Feynman graphs:
\begin{equation}
  \label{eq:amp-G}
  G^v_\amp = \phi_R(X^v_\amp) = \frac{G^v}{\prod_{e \sim v}
    \sqrt{G^e}}\,. 
\end{equation}
In physics literature, $G^v_\amp$ is often referred to as the \textit{amputated Green's function for the vertex
  $v$}. In more mathematical literature, these functions are called the \textit{invariant charges}.

Let us shortly motivate the above definition by showing how the amputated Green's functions appear in physics
computations. For example, consider any particle interaction process that is described by the QFT. Let $v$ be
the residue of the respective Feynman graphs. Then, one can easily show that the corresponding Feynman amplitude
(including all quantum corrections) is given by $\prod_{e \sim v} (G^e )^{-1 \slash 2}$ times a polynomial of
the amputated Green's functions. For example, the righthandsides of Eqs.~(\ref{eq:ex3}) and (\ref{eq:ex4}) read
\begin{equation*}
  \SetScale{0.3} \frac{G_\amp^{\begin{picture}(9,0)(0,0) \Photon(0,18)(16,18){2}{2.5} \Vertex(16,18){1.7}
        \ArrowLine(16,18)(30,32)
        \ArrowLine(30,4)(16,18) \end{picture}}}{\sqrt{G^{\begin{picture}(6,0)(0,0) 
        \Photon(0,18)(20,18){2}{3.5}\end{picture}}} G^{\begin{picture}(6,0)(0,0)
        \ArrowLine(0,18)(20,18)\end{picture}}} \qquad \text{and} \qquad \SetScale{0.3}
  \frac{G_\amp^{\begin{picture}(9,0)(0,0) \ArrowLine(16,18)(2,32) \ArrowLine(2,4)(16,18)
        \Vertex(16,18){1.7} \ArrowLine(16,18)(30,32) \ArrowLine(30,4)(16,18) \end{picture}} +
    \left(G_\amp^{\begin{picture}(9,0)(0,0) \Photon(0,18)(16,18){2}{2.5} \Vertex(16,18){1.7}
        \ArrowLine(16,18)(30,32)
        \ArrowLine(30,4)(16,18) \end{picture}} \right)^2}{\left(G^{\begin{picture}(6,0)(0,0) 
        \ArrowLine(0,18)(20,18)\end{picture}}\right)^2}\,.
\end{equation*}

Secondly, it is common in particle physics to redefine the coupling constants of the theory as
$\tilde{g}_k = g_k G^{v_k}_\amp$. This makes them dependent on the energy scale and has the advantage that a
perturbative expansion for the Green's functions in terms of $\tilde{g}_k$ and the logarithmic energy scale $L$
may behave much better. This is an alternative way to approximate the Green's functions accurately without using
the log-expansion. The $\tilde{g}_k$ are called the \textit{running coupling parameters}. In our case, there is
only one coupling constant and we redefined it such that $\alpha = g^l$ counts the number of loops in a Feynman
graph. Hence, we must also redefine the respective invariant charge as~\footnote{Here, $v$ is the vertex residue
  that corresponds to the interaction term in the Lagrangian.}
\begin{equation}
  \label{eq:Q}
  X_Q = \left(X^v_\amp \right)^l\,,\qquad Q(\alpha,L) = \phi_R(X_Q) = G^v_\amp(\alpha,L)^l \,.
\end{equation}
Then, the running coupling parameter is $\tilde{\alpha} = \alpha Q$. At the renormalization scale ($L = 0$), one
has $\tilde{\alpha} = \alpha$, because $Q(\alpha,0) = 1$.

Now, the anomalous dimension $\gamma^r$ is defined as the negative of the $L$-linear part of the Green's function
$G^r$,
\begin{equation}
  \label{eq:gamma}
  \gamma^r(\alpha) = \sum_{k \geq 0}\gamma^r_k \alpha^{k+1} = - \frac{\partial G^r(\alpha,L)}{\partial L}
  \bigg|_{L = 0}\,.
\end{equation}
Similarly, the $\beta$-function (in our case, there is only one) is the $L$-linear part of the invariant charge,
\begin{equation}
  \label{eq:beta}
  \beta(\alpha) = \sum_{k \geq 0}\beta_k \alpha^{k+1} = \frac{\partial Q(\alpha,L)}{\partial L} \bigg|_{L =
    0} = l 
  \left(
    \frac{1}{2} \sum_{e \sim v} \gamma^e(\alpha) - \gamma^v(\alpha)
  \right)\,.
\end{equation}
Indeed, it follows from Eq.~(\ref{eq:amp-G}) that it is just a linear combination of the various anomalous
dimensions. For example, $l = 2$ in QED:
\begin{equation}
  \label{eq:QED-beta}
  Q(\alpha,L) = \SetScale{0.3} G_\amp^{\begin{picture}(9,0)(0,0) \Photon(0,18)(16,18){2}{2.5}
      \Vertex(16,18){1.7} \ArrowLine(16,18)(30,32)
      \ArrowLine(30,4)(16,18) \end{picture}}(\alpha,L)^2\,,\qquad   \beta = \SetScale{0.3}
  2\gamma^{\begin{picture}(6,0)(0,3) \ArrowLine(0,18)(20,18)\end{picture}} + \gamma^{\begin{picture}(6,0)(0,3)
      \Photon(0,18)(20,18){2}{3.5}\end{picture}} - 2 \gamma^{\begin{picture}(9,0)(0,3)
      \Photon(0,18)(16,18){2}{2.5} \Vertex(16,18){1.7} \ArrowLine(16,18)(30,32)
      \ArrowLine(30,4)(16,18) \end{picture}}\,. 
\end{equation}

Two remarks are in order: First, the notion of an anomalous dimension $\gamma^r$ for vertex type residues is
quite unusual in particle physics. Indeed, it is redundant, since $\gamma^v$ can be expressed by the
$\beta$-function and the various propagator anomalous dimensions $\gamma^e$ via Eq.~(\ref{eq:beta}). In our case
however, the definition in Eq.~(\ref{eq:gamma}) including vertex type residues is very convenient because the
formulas for the log-expansions of the Green's functions $H^r$ generalize to \textit{all kinds of residues}.
Secondly, the minus sign in Eq.~(\ref{eq:gamma}) is chosen such that it cancels the sign in
Eq.~(\ref{eq:X-propagator}) for propagator type residues $r = e$. In this case, the anomalous dimension
$\gamma^e$ is given by the $L$-linear part of the Feynman amplitude that corresponds to the sum of all
respective 1PI propagator graphs.

The functions $\gamma^r$ and $\beta$ can be computed perturbatively. Neglecting terms of order
$\mathcal{O}(\alpha^{n+1})$ in Eqs.~(\ref{eq:gamma}) and (\ref{eq:beta}) gives the so-called $n$-loop anomalous
dimensions and $n$-loop $\beta$-function. These approximations can be obtained by only evaluating Feynman graphs
with at most $n$ loops. It turns out that the N$^{(n)}$LL approximation of the Green's functions $H^r$ only
depend on the $(n+1)$-loop anomalous dimensions $\gamma^r$ and $\beta$-function. We show this explicitly in the
next section.
 
\subsection{Log-Expansions}
\label{sec:log-expansion}

We start with the following formula for the Green's functions: 
\begin{equation}
  \label{eq:MAIN}
  \frac{\partial \log G^r}{\partial L} = - \gamma^r(\alpha Q)\,.
\end{equation}
A detailed derivation from combinatorial Dyson-Schwinger equations is collected in Section~\ref{sec:review} (for
a short proof, see Section~\ref{sec:derivation-1}). Taking a linear combination of the above relations for
different residues $r$ implies an ordinary differential equation for the invariant charge, i.e.
\begin{equation}
  \label{eq:dQdL}
  \frac{\partial \log Q}{\partial L} = \beta(\alpha Q).
\end{equation}
Together with the initial conditions
\begin{equation}
  \label{eq:initial}
  G^r(\alpha,0) = \phi_R(r) = 1\,, \qquad Q(\alpha,0) = 1\,,
\end{equation}
Eqs.~(\ref{eq:MAIN}) and (\ref{eq:dQdL}) determine the full $L$-dependence of the Green's functions
$G^r$~\footnote{Note that this only requires to know the anomalous dimensions $\gamma^r$ and $\beta$-function of
  the QFT, i.e. to say the $L$-linear parts of the Green's functions.}.

In particle physics, the above relations are usually written in terms of the ($L$-dependent) running coupling
parameter $\tilde{\alpha} = \alpha Q$. Then, Eq.~(\ref{eq:dQdL}) corresponds to the \textit{renormalization
  group equation} (RGE)~\footnote{The usual definition of the $\beta$-function in the literature slightly
  differs from ours:
  \begin{equation*}
    \beta_\text{lit}(\tilde{\alpha}) = \tilde{\alpha} \beta(\tilde{\alpha})\,.  
  \end{equation*}
  E.g., the QED 1-loop $\beta$-function is $\beta_\text{lit}(\tilde{\alpha}) = \beta_0 \tilde{\alpha}^2$. The
  notations in this paper are consistent, such that our 1-loop $\beta$-function is \textit{linear} in the
  (loop-counting) coupling parameter.} and Eq.~(\ref{eq:MAIN}) represents the Callan-Symanzik equation for the
Green's function $\tilde{G}^r$ with $\tilde{G}^r(\tilde{\alpha},L) = G^r(\alpha,L)$:
\begin{equation*}
  \frac{\partial
    \tilde{\alpha}}{\partial L} = \tilde{\alpha} \beta(\tilde{\alpha})\,,\qquad   \left(
    \frac{\partial}{\partial L} + \tilde{\alpha} \beta(\tilde{\alpha}) \frac{\partial}{\partial \tilde{\alpha}}
    + \gamma^r(\tilde{\alpha}) \right) \tilde{G}^r(\tilde{\alpha},L) = 0 \,.
\end{equation*}

The derivation of the log-expansions for the Green's functions requires another differential equation for the
invariant charge $Q$. Therefore, one integrates Eq.~(\ref{eq:dQdL}) using separation of variables. With the
above definition of the running coupling parameter, this leads to
\begin{equation*}
  \int^{\tilde{\alpha}}_\alpha \frac{\mathrm{d} x}{x \beta(x)} = L\,.
\end{equation*}
Here, the integration constant is well chosen such that $\tilde{\alpha} = \alpha$ for $L = 0$. Since we cannot
solve this for $Q$, it is more convenient to take the derivative with respect to~$\alpha$. This results in
\begin{equation}
  \label{eq:dQda}
  1 + \alpha \frac{\partial \log Q}{\partial \alpha} = \frac{\beta(\alpha Q)}{\beta(\alpha)}\,.  
\end{equation}

With these considerations, we can now give the main equations that determine the log-expansions for the Green's
functions. One therefore defines a log-expansion for the invariant charge~\footnote{Accordingly, the
  log-expansion for the running coupling is given by $\tilde{\alpha}(\alpha,z) = \alpha R(\alpha,z)$.},
\begin{equation}
  \label{eq:R}
  Q(\alpha,L) = R(\alpha,\alpha L)\,, \qquad R(\alpha,z) = R_0(z) - \sum_{j\geq 1} \alpha^j R_j(z)\,,
\end{equation}
and a change of variables in Eqs.~(\ref{eq:MAIN} - \ref{eq:dQda}) finally results in
\begin{gather}
  \frac{\partial \log H^r}{\partial z} = - \frac{\gamma^r(\alpha R)}{\alpha}\,, \qquad H^r(\alpha,0) =
  1\,, \label{eq:logH}\\[5pt] 
  \frac{\partial \log R}{\partial z}   = \frac{\beta(\alpha R)}{\alpha}\,, \qquad R(\alpha,0) =
  1\,, \label{eq:logR}\\[5pt] 1 + \alpha \frac{\partial \log R}{\partial \alpha} - \frac{\beta(\alpha
    R)}{\beta(\alpha)} \left( 1 - \frac{z \beta(\alpha)}{\alpha} \right) = 0\,. \label{eq:rek}
\end{gather}
Now, one uses Eqs.~(\ref{eq:logR}) and (\ref{eq:rek}) in order to find the N$^{(n)}$LL approximation of the
invariant charge~\footnote{This is equivalent to solving the RGE and implies a log-expansion for the running
  coupling parameter $\tilde{\alpha}$, see e.g. \cite{Chetyrkin:2000yt}.}. The obtained result can then be used
in Eq.~(\ref{eq:logH}) to deduce the N$^{(n)}$LL approximation of the Green's functions $H^r$.

The first step is simply achieved by a Taylor series expansion of Eq.~(\ref{eq:rek}) in $\alpha$. The $0$th
order reads
\begin{equation}
  \label{eq:R-ll}
  R_0(z) = \frac{1}{1 - \beta_0 z}\,,
\end{equation}
which gives the LLO of the invariant charge. Now, consider all terms of order $\alpha^n$ in Eq.~(\ref{eq:rek})
with $n\geq 1$ and note that they do not contain any of the functions $R_N(z)$ with $N > n$. Hence, collecting
all terms including the function $R_n(z)$, one finds
\begin{equation*}
  \frac{1 - n}{R_0(z)} R_n(z) = \ldots\,,
\end{equation*}
where the rhs only consists of the functions $R_0,\ldots, R_{n - 1}$. This is a nice recursive formula for the
various log orders of the invariant charge, except for $n = 1$. Two remarks are in order: First, the
$\alpha$-linear terms in Eq.~(\ref{eq:rek}) vanish completely, because an $\alpha$-derivative was taken in order
to obtain Eq.~(\ref{eq:dQda}). However, $R_1$ can be found from the $\alpha$-linear part of Eq.~(\ref{eq:logR}),
which reads
\begin{equation*}
  \frac{\mathrm{d} R_1}{\mathrm{d} z} = 2 \beta_0 R_0 R_1 - \beta_1 R_0^3\,. 
\end{equation*}
The solution of this ordinary first-order differential equation with the respective initial condition
$R_1(0) = 0$ is given by
\begin{equation}
  \label{eq:R-nll}
  R_1(z) = -\frac{\beta_1}{\beta_0} R_0(z)^2\log R_0(z)\,.
\end{equation}
Secondly, an explicit form of the recursive relation for $R_n$ is not required, because Eq.~(\ref{eq:rek}) with
cleared fractions can be given directly to a computer algebra program in order to obtain the various log orders.
For example, the $\alpha^2$ order in Eq.~(\ref{eq:rek}) reads
\begin{equation}
  \label{eq:R-nnll}
  R_2(z) = R_0(z)^3
  \left(
    - \beta_2 z + \frac{\beta_1^2}{\beta_0^2} 
    \left(
      \beta_0 z - \log R_0(z) - \log^2 R_0(z)
    \right)
  \right)\,,
\end{equation}
when Eqs.~(\ref{eq:R-ll}) and (\ref{eq:R-nll}) are used.

The second step (the derivation of the log-expansion for the Green's functions $H^r$) is a bit more involved,
because one has to integrate $\gamma^r(\alpha R)$ with respect to $z$. Indeed, from Eq.~(\ref{eq:logH}),
\begin{equation}
  \label{eq:exp-greens-function}
  H^r(\alpha,z) = \exp 
  \left(
    - \int_0^z \frac{\gamma^r(\alpha R(\alpha,z'))}{\alpha}\, \mathrm{d}z'
  \right)\,.
\end{equation}
In order to obtain the N$^{(n)}$LL approximation of $H^r$, one needs to expand the exponent up to order
$\alpha^{n+1}$. As promised earlier, this requires knowledge of the $(n+1)$-loop anomalous dimension $\gamma^r$
and $\beta$-function. The computation is quite technical but can always be done with the help of a computer
algebra program. Here, we contentedly give the NNLL approximation, i.e.
\begin{align}
\label{eq:NNLL-H}
  H^r(\alpha,z) = R_0^{\,}(z)^{-\gamma^r_0 \slash \beta^{\,}_0}\, 
  \Bigg[ & \qquad 1 \qquad  + \qquad \frac{\alpha R_0^{\,}(z)}{\beta_0^2}
           \big(
           \gamma^r_0 \beta^{\,}_1 (1 - \log R_0^{\,}(z)) - \gamma^r_1 \beta^{\,}_0 
           \big)\nonumber\\[5pt]
  +\, \frac{\alpha^2 R_0^{\,}(z)^2}{2\beta_0^4}\, \times\,
          &\bigg(
                      \left(\beta^{\,}_1 \gamma^r_0 - \beta^{\,}_0 \gamma^r_1 \right)^2 + \beta^{\,}_0 \beta_1^2
           \gamma^r_0 - \beta_0^2 \beta^{\,}_2 \gamma^r_0 + \beta_0^2 \beta^{\,}_1 \gamma^r_1 - \beta_0^3
            \gamma^r _2 
           \nonumber\\ 
         &\;+ 2  
           \left(
           \beta_0^2 \beta^{\,}_2 \gamma^r_0 - \beta^{\,}_0 \beta_1^2 \gamma^r_0
           \right) R_0^{\,}(z)^{-1} \nonumber\\[5pt]
         &\;- 2 
           \left(
           \beta_1^2 (\gamma^r_0 )^2 - \beta^{\,}_0 \beta^{\,}_1 \gamma^r_0 \gamma^r_1 + \beta_0^2
           \beta^{\,}_1 \gamma^r_1
           \right) \log R_0^{\,}(z)
           \nonumber\\ 
         &\;+
           \left(
           \beta_1^2 (\gamma^r_0)^2 - \beta^{\,}_0 \beta_1^2 \gamma^r_0
           \right) \log^2 R_0^{\,}(z) \bigg) \qquad+ \mathcal{O}(\alpha^3) \qquad
           \Bigg]\,.
\end{align}

\subsection{Examples: Truncated Propagator Green's Functions and QED}
\label{sec:exampl-qed-trunc}

Let us give a short example of the above formula~(\ref{eq:NNLL-H}). In QED, the Ward-Takahashi identities state
that the Green's functions for the fermion propagator
$\SetScale{0.3}G^{\begin{picture}(6,0)(0,0) \ArrowLine(0,18)(20,18)\end{picture}}$ and the 3-point vertex
$ \SetScale{0.3} G^{\begin{picture}(9,0)(0,0) \Photon(0,18)(16,18){2}{2.5} \Vertex(16,18){1.7}
    \ArrowLine(16,18)(30,32) \ArrowLine(30,4)(16,18) \end{picture}}$
coincide. Therefore, the invariant charge in Eq.~(\ref{eq:Q}) is given by the inverse of the photon propagator
Green's function. Furthermore, the $\beta$-function equals the respective anomalous dimension, see
Eq.~(\ref{eq:QED-beta}):
\begin{equation*}
  \SetScale{0.3}
  Q = \frac{1}{G^{\begin{picture}(6,0)(0,0) \Photon(0,18)(20,18){2}{3.5}\end{picture}}} \qquad \text{and} \qquad \beta = \gamma^{\begin{picture}(6,0)(0,3) \Photon(0,18)(20,18){2}{3.5}\end{picture}} \,.
\end{equation*}

The above relation is a special case of a broader class of QFT toy models that describe only one particle
(edge-type $e$ in Feynman graphs), where the invariant charge and $\beta$-function are given by 
\begin{equation}
  \label{eq:beta-s}
  Q = \left(G^e \right)^{-s} \qquad \text{and} \qquad \beta = s \gamma^e
\end{equation}
for integers $s$. Examples are $s = 1$ in QED, as well as $s = 2$ ($s = 3$) for a truncated Yukawa theory
($\phi^3$-theory), where all vertex divergences are neglected. In all these cases, there is only one propagator
Green's function $G^e$ with log-expansion $H^{e(s)}$ and it is given by $R^{-1 \slash s}$~\footnote{Note that
  this is consistent with Eqs.~(\ref{eq:logR}, \ref{eq:exp-greens-function} and \ref{eq:beta-s}).}. The NNLL
approximation can be obtained in two ways: Either by inserting the relation between the $\beta$-function and the
anomalous dimension into Eq.~(\ref{eq:NNLL-H}); or by using the explicit expression for the log-expansion of the
invariant charge in Eqs.~(\ref{eq:R}, \ref{eq:R-ll} - \ref{eq:R-nnll}). In both cases, we obtain~\footnote{In
  \cite{Yeats}, equivalent expressions have been derived using Chord diagrams. These formulas contain the
  symbols $a_{i,j}$, which are related to the anomalous dimension $\gamma^e$ as follows:
  \begin{gather*}
    \gamma^{e(1)}_0 = - a_{1,0}\,,\qquad \gamma^{e(1)}_1 = - a_{2,0},\,\qquad \gamma^{e(1)}_2 =
    - (a_{3,0} + a_{2,1} a_{1,0})\,,\qquad \gamma^{e(2)}_0 = - a_{1,0} \\
    \gamma^{e(2)}_1 = - (a_{2,0} + a_{1,1}a_{1,0}),\,\qquad \gamma^{e(2)}_2 = - (a_{3,0} + a_{1,1}a_{2,0} +
    a_{1,1}^2 a^{\,}_{1,0} + 3 a^{\,}_{1,2}a_{1,0}^2 + 3 a_{2,1} a_{1,0})\,.\\[-28pt]
  \end{gather*}
}
\begin{align}
  \label{eq:H_s}
  H^{e(s)}(\alpha,z) =\,& R_0^{\,}(z)^{-1\slash s} 
  \Bigg[ \quad 1 \quad - \quad \alpha \, \frac{\beta_1 R_0^{\,}(z)}{s \beta_0}\, \log R_0^{\,}(z) \quad + \quad
          \alpha^2\,  
          \frac{ \beta_1^2 R_0^{\,}(z)^2}{s \beta_0^2} \times \nonumber\\
        & \quad\times
          \left(
          \beta_0 z 
          \left(
          1 - \frac{\beta_0 \beta_2}{\beta_1^2}
          \right) - \log R_0^{\,}(z) + \frac{1 - s}{2s} \log^2 R_0^{\,}(z)
          \right) + \mathcal{O}(\alpha^3) \Bigg]\,.
\end{align}

\section{Mathematical Background}
\label{sec:review}

This section collects the essential properties of Feynman graphs and renormalized Feynman rules (presented in a
mathematical framework) that lead to the Callan-Symanzik equation~(\ref{eq:MAIN}). The derivation of this
relation is given in Section~\ref{sec:derivation-1} and requires two main ingredients: The first one is the well
known \textit{exponential formula} for $\phi_R$, see Eq.~(\ref{eq:exp-formula}). We motivate this property of
renormalized Feynman rules and give a short introduction to the Hopf algebra of Feynman graphs in
Section~\ref{sec:hopf}. The second ingredient is an (also well known) expression for the co-product of the
infinite sums $X^r$ in Eqs.~(\ref{eq:X-propagator}) and (\ref{eq:X-vertex}), see Eq.~(\ref{eq:Delta-Xr}). We
review in Section~\ref{sec:graph-insertion}, how this relation is obtained from fixed-point equations for $X^r$,
i.e. so-called combinatorial Dyson-Schwinger equations (DSE's). In the literature, the latter are written down
in full detail only for those sums $X^r$ with $\phi_R(r) = 1$~\footnote{Hence, $|r| = 2$ or $r = v$ is the
  vertex that corresponds to the interaction term in the Lagrangian.}, which is the reason why we restrict our
results to that case. For presentation purposes, we consider QFT's with only one interaction term in the
Lagrangian (we explain the redefinition of the coupling constant in Section~\ref{sec:grading-mathcalh}).
However, DSE's also exist in theories with more than one coupling parameter --- the generalization to that case
is given in Section~\ref{sec:gener-results}.

\subsection{The Grading of $\mathcal{H}$}
\label{sec:grading-mathcalh}

For QFT's with only one interaction term in the Lagrangian, one redefines the coupling constant as
$\alpha = g^l$, which implies that a $K$-loop graph maps to a term proportional to $\alpha^K$ under renormalized
Feynman rules. Let us show this now.

First, \textit{Euler's formula} states that for a graph with $K$ loops, $V$ vertices and $E$ internal edges, the
following identity holds:
\begin{equation*}
  K + V - E = 1\,.
\end{equation*}
Secondly, let $|v|$ be the degree of the vertex $v$~\footnote{i.e. the number of edges which are incident to
  $v$} that corresponds to the interaction term in the Lagrangian. For example, $|v| = 3$ in QED and $|v| = 4$
in $\phi^4$-theory. Then, a Feynman graph with residue $r$ has $|r|$ external edges, hence
\begin{equation*}
  V\cdot |v| = 2E + |r|\,.
\end{equation*}
Together with Euler's formula and the definition of $l:= 2\slash (|v| - 2)$, one finally finds
\begin{equation}
  \label{eq:V-L}
  V = l K +
  \begin{cases}
    0,& \text{for } |r| = 2 \\
    1, & \text{for } |r| = |v|
  \end{cases}\,.
\end{equation}
One concludes that with the correct normalization of renormalized Feynman rules, a graph with $K$ loops and $V$
vertices maps to a term that is proportional to $g^{l K} = \alpha^K$.

\subsection{Feynman Rules from a Hopf Algebraic Point of View}
\label{sec:hopf}

This section represents a minimal introduction to the Hopf algebra of Feynman graphs
\cite{Kreimer:1997dp,Connes:1999yr}. As a motivation, consider a typical observation during the renormalization
of Feynman amplitudes:
\begin{equation}
  \label{eq:feynmanrules-prop}
  \SetScale{0.4}
  \phi_R^{L_1 + L_2} 
  \left(
    \begin{picture}(32,10)(0,15.5)
      \ArrowLine(0,36)(80,36) \PhotonArc(40,36)(20,0,70){2.2}{3.5} \PhotonArc(40,36)(20,110,180){2.2}{3.5}
      \LongArrowArcn(40,55)(7,70,250) \LongArrowArcn(40,55)(7,250,430)
      \Vertex(33,55){1.7} \Vertex(47,55){1.7} \Vertex(20,36){1.7} \Vertex(60,36){1.7}
    \end{picture}
  \right) = \phi_R^{L_1} 
  \left(
    \begin{picture}(32,10)(0,15.5)
      \ArrowLine(0,36)(80,36) \PhotonArc(40,36)(20,0,70){2.2}{3.5} \PhotonArc(40,36)(20,110,180){2.2}{3.5}
      \LongArrowArcn(40,55)(7,70,250) \LongArrowArcn(40,55)(7,250,430)
      \Vertex(33,55){1.7} \Vertex(47,55){1.7} \Vertex(20,36){1.7} \Vertex(60,36){1.7}
    \end{picture}
  \right) + \phi_R^{L_2} 
  \left(
    \begin{picture}(32,10)(0,15.5)
      \ArrowLine(0,36)(80,36) \PhotonArc(40,36)(20,0,70){2.2}{3.5} \PhotonArc(40,36)(20,110,180){2.2}{3.5}
      \LongArrowArcn(40,55)(7,70,250) \LongArrowArcn(40,55)(7,250,430)
      \Vertex(33,55){1.7} \Vertex(47,55){1.7} \Vertex(20,36){1.7} \Vertex(60,36){1.7}
    \end{picture}
  \right) + \phi_R^{L_1} 
  \left(
    \SetScale{0.35}
    \begin{picture}(28,10)(0,10)
      \Photon(0,36)(20,36){3.3}{2.5} \Vertex(20,36){1.7} 
      \Photon(60,36)(80,36){3.3}{2.5} \Vertex(60,36){1.7}
      \LongArrowArcn(40,36)(20,80,260) \LongArrowArcn(40,36)(20,260,440)
    \end{picture}
  \right) \phi_R^{L_2} 
  \left( 
    \SetScale{0.4}
    \begin{picture}(32,10)(0,15.5)
      \ArrowLine(0,36)(80,36) \PhotonArc(40,36)(20,0,180){2.3}{7.5} \Vertex(20,36){1.7} \Vertex(60,36){1.7}
    \end{picture}
  \right)\,.
\end{equation}
Here, and only within this section, we write renormalized Feynman rules $\phi_R^{L_i}$ with an upper index
$L_i$. It indicates the energy scale $L_i = \log S_i \slash S_0$ at which Feynman graphs are evaluated. The goal
of this section is to generalize the above observation (to the exponential formula). This can be achieved in a
very elegant way, once we introduced the Hopf algebra of Feynman graphs.

Let us start with the algebra $\mathcal{H}$, i.e. the vector space $\mathcal{H}$ equipped with an associative
product $m : \mathcal{H} \otimes \mathcal{H} \to \mathcal{H}$ (the disjoint union) and a unit element
$\mathbb{I}$ (the empty graph). In a more mathematical notation,
\begin{align*}
  \text{associativity:} \;\,\quad m \circ (m \otimes \id) &= m \circ (\id \otimes m) \,, \\
  \text{unit element:}\, \qquad m(X \otimes \mathbb{I}) \;\; &= \;\;
                           m(\mathbb{I} \otimes X) = X \qquad \forall X \in \mathcal{H}\,,
\end{align*}
where $\id$ denotes the identity map on $\mathcal{H}$.

In a first step, a \textit{co-product} $\Delta$ and a \textit{co-unit} $\hat{\mathbb{I}}$ are defined, such that
$\mathcal{H}$ extends to a \textit{bi-algebra}. The important tool for our purposes is the co-product. It is a
linear map $\Delta: \mathcal{H} \to \mathcal{H} \otimes \mathcal{H}$, which decomposes a graph into its
sub-graphs in a certain way. The explicit definition of its action on 1PI Feynman graphs is given by
\begin{equation}
  \label{eq:co-product}
  \Delta \Gamma = \Gamma \otimes \mathbb{I} + \mathbb{I} \otimes \Gamma + \sum_{\gamma \subsetneq \Gamma} \gamma
  \otimes \Gamma \slash \gamma \,,
\end{equation}
where the sum extends over (disjoint unions of) divergent sub-graphs $\gamma$ of $\Gamma$. Furthermore,
$\Gamma \slash \gamma$ corresponds to $\Gamma$ but with the sub-graphs $\gamma$ shrinked to a point. For example,
\begin{equation}
\SetScale{0.4}
  \Delta \left(
  \begin{picture}(32,10)(0,15.5)
    \ArrowLine(0,36)(80,36) \PhotonArc(40,36)(20,0,70){2.2}{3.5} \PhotonArc(40,36)(20,110,180){2.2}{3.5}
    \LongArrowArcn(40,55)(7,70,250) \LongArrowArcn(40,55)(7,250,430)
    \Vertex(33,55){1.7} \Vertex(47,55){1.7} \Vertex(20,36){1.7} \Vertex(60,36){1.7}
  \end{picture} \right)
  = 
  \begin{picture}(32,10)(0,15.5)
    \ArrowLine(0,36)(80,36) \PhotonArc(40,36)(20,0,70){2.2}{3.5} \PhotonArc(40,36)(20,110,180){2.2}{3.5}
    \LongArrowArcn(40,55)(7,70,250) \LongArrowArcn(40,55)(7,250,430)
    \Vertex(33,55){1.7} \Vertex(47,55){1.7} \Vertex(20,36){1.7} \Vertex(60,36){1.7}
  \end{picture}
  \otimes \mathbb{I} + \mathbb{I} \otimes
  \begin{picture}(32,10)(0,15.5)
    \ArrowLine(0,36)(80,36) \PhotonArc(40,36)(20,0,70){2.2}{3.5} \PhotonArc(40,36)(20,110,180){2.2}{3.5}
    \LongArrowArcn(40,55)(7,70,250) \LongArrowArcn(40,55)(7,250,430)
    \Vertex(33,55){1.7} \Vertex(47,55){1.7} \Vertex(20,36){1.7} \Vertex(60,36){1.7}
  \end{picture}
  + \SetScale{0.35}
  \begin{picture}(28,10)(0,10)
    \Photon(0,36)(20,36){3.3}{2.5} \Vertex(20,36){1.7} 
    \Photon(60,36)(80,36){3.3}{2.5} \Vertex(60,36){1.7}
    \LongArrowArcn(40,36)(20,80,260) \LongArrowArcn(40,36)(20,260,440)
  \end{picture}
  \otimes \SetScale{0.4}
  \begin{picture}(32,10)(0,15.5)
    \ArrowLine(0,36)(80,36) \PhotonArc(40,36)(20,0,180){2.3}{7.5} \Vertex(20,36){1.7} \Vertex(60,36){1.7}
  \end{picture}\,.
\end{equation}
On a disjoint union of Feynman graphs, $\Delta$ is defined such that it is compatible with the product (a
requirement for bi-algebras),
\begin{equation}
  \label{eq:co-product2}
  \Delta (XY) = \sum X'Y' \otimes X'' Y''\,.
\end{equation}
Here, Sweedler's notation for the co-product is used, e.g. $\Delta X = \sum X' \otimes X''$. The co-unit
$\I : \mathcal{H} \to \mathbb{K}$ maps an element $X \in \mathcal{H}$ to the coefficient of $\mathbb{I}$ in $X$.
Indeed, it is easy to show that the co-product is co-associative and that $\I$ is the co-unit element for
$\Delta$,
\begin{align*}
  \text{co-associativity:} \quad (\Delta \otimes \id)\circ\Delta &= (\id \otimes \Delta)\circ\Delta \,, \\
  \text{co-unit element:}\quad\:\, (\I \otimes \id)\circ\Delta &=
                                                               (\id \otimes \I)\circ\Delta \simeq \id \,.
\end{align*}
Hence, $(\mathcal{H},m,\mathbb{I},\Delta,\I)$ forms a bi-algebra.

Secondly, one defines an \textit{antipode} to further extend $\mathcal{H}$ to a Hopf algebra. It is the linear
map $S:\mathcal{H} \to \mathcal{H}$, which is recursively given by
\begin{equation*}
  m \circ(S \otimes \id)\circ \Delta =  m \circ (\id \otimes S)\circ \Delta = \mathbb{I} \circ\I\,, \qquad S\circ m = m\circ (S
  \otimes S)\,. 
\end{equation*}
All in all, $(\mathcal{H},m,\mathbb{I},\Delta,\I,S)$ forms a Hopf algebra, called the Hopf algebra of Feynman
graphs.

With the above defined structures, the vector space $\Hom(\mathcal{H},\mathcal{A})$ of algebra homomorphisms
extends to a group $G_\mathcal{H}^\mathcal{A}$, which is called the \textit{convolution- or character-group}
\cite{Connes:2000fe}. For example, renormalized Feynman rules are a \textit{character},
$\phi_R \in G_\mathcal{H}^\mathcal{A}$. The conjunction in $G_\mathcal{H}^\mathcal{A}$ is the convolution
product $\star$: Let $\phi, \psi \in G_\mathcal{H}^\mathcal{A}$. Then,
\begin{equation}
  \label{eq:star}
  \phi \star \psi := m_\mathcal{A} \circ (\phi \otimes \psi) \circ \Delta 
\end{equation}
is an algebra homomorphism, hence $\phi \star \psi \in G_\mathcal{H}^\mathcal{A}$. The convolution is
associative because the co-product is co-associative and the product $m_\mathcal{A}$ in $\mathcal{A}[\alpha,L]$
(multiplication) is associative. The unit element in $G_\mathcal{H}^\mathcal{A}$ is the linear map
$e = 1 \circ \I$ and the inverse of a character $\phi$ is given by $\phi \circ S$.

The observation in Eq.~(\ref{eq:feynmanrules-prop}) can now be generalized in a beautiful way: Renormalized
Feynman rules satisfy the property~\footnote{This is long known but never really written down in its full
  beauty. See e.g. \cite{Brown:2011pj}. Note that Eq.~(\ref{eq:feynmanrules-prop}) is just a special case of
  Eq.~(\ref{eq:main-phi}).}
\begin{equation}
  \label{eq:main-phi}
  \phi_R^{L_1 + L_2} = \phi_R^{L_1} \star \phi_R^{L_2}\,.
\end{equation}
It follows that the vector space of maps $\phi_R^L$, equipped with the convolution product, forms a group. This
is the so-called \textit{renormalization group} and it is isomorphic to the additive group. Hence, there exists
a linear map $\sigma \in G_\mathcal{H}^\mathcal{A}$, such that
\begin{equation}
  \label{eq:exp-formula}
  \phi_R = \exp^\star (L \sigma) = e + L \sigma + \frac{L^2}{2} \sigma \star \sigma + \ldots
\end{equation}
This is the \textit{exponential formula}. Here, we dropped the upper index in $\phi_R$ again, in order to adapt
the notation to the rest of this paper. We can give a meaning to the character $\sigma$: It maps a (disjoint
union of) Feynman graph(s) to the $L$-linear part of the corresponding Feynman amplitude,
\begin{align*}
  \sigma : \mathcal{H} & \to \; \mathcal{A}\\
  X & \mapsto \sigma(X) = \frac{\partial \phi_R(X)}{\partial L} \Big|_{L = 0}\,.
\end{align*}
In particular, the infinite sums $X^r$ in Eqs.~(\ref{eq:X-propagator}) and (\ref{eq:X-vertex}) are mapped to the
negative of the anomalous dimension (see Eq.~(\ref{eq:gamma})),
\begin{equation}
  \label{eq:sigma-anom}
  \sigma (X^r) = -\gamma^r(\alpha) = - \sum_{k\geq 0} \gamma^r_k \alpha^{k+1}\,.
\end{equation}

From now on, we do not need the co-unit and antipode of the Hopf algebra anymore. These structures were only
required for a proper definition of the convolution group $G_\mathcal{H}^\mathcal{A}$. The proof of
Eq.~(\ref{eq:MAIN}) only uses the co-product and the exponential formula in Eq.~(\ref{eq:exp-formula}).

\subsection{Graph Insertion and Dyson-Schwinger Equations}
\label{sec:graph-insertion}

In this section, we motivate the following formula for the co-product~\footnote{This formula is well known in
  the literature, see e.g. \cite{Yeats:2008zy,vanSuijlekom:2008kv,Borinsky:2014xwa,Kissler:2016gxn}, or in a
  slightly different context, also \cite{Bergbauer:2005fb,Foissy_faadi}.} of the infinite sums $X^r$ in
Eqs.~(\ref{eq:X-propagator}) and (\ref{eq:X-vertex})~\footnote{These map to the Green's functions $G^r$ under
  renormalized Feynman rules.}:
\begin{equation}
  \label{eq:Delta-Xr}
  \Delta X^r = \sum_{k\geq 0} X^r X_Q^k \otimes X^r \big|_{k}\,.
\end{equation}
Here, $X_Q \in \mathcal{H}$ is the combinatorial invariant charge ($\phi_R(X_Q) = Q$), see Eq.~(\ref{eq:Q}). For
example in QED:
\begin{equation*}
  \SetScale{0.3} X_Q = \frac{\left(X^{\begin{picture}(9,0)(0,0)
          \Photon(0,18)(16,18){2}{2.5} \Vertex(16,18){1.7} \ArrowLine(16,18)(30,32)
          \ArrowLine(30,4)(16,18) \end{picture}} \right)^2}{X^{\begin{picture}(6,0)(0,0)
        \Photon(0,18)(20,18){2}{3.5}\end{picture}} \left(X^{\begin{picture}(6,0)(0,0)
          \ArrowLine(0,18)(20,18)\end{picture}}\right)^2}\,.
\end{equation*}
Furthermore, $|_k$ denotes the projection onto $\mathcal{H}_k$, hence $\phi_R(X|_k) \propto \alpha^k$ for each
$X \in \mathcal{H}$.

Let us start by introducing the notion of \textit{graph insertion}. Therefore, consider for example the QED
Feynman graphs
\begin{equation}
  \label{eq:ex5}
  \SetScale{0.5}
  \begin{picture}(40,25)(0,10)
    \Photon(0,36)(32,36){3.3}{3.5} \Vertex(32,36){1.7}
    \Line(80,72)(32,36) \ArrowLine(64,12)(32,36)
    \Line(80,0)(32,36) \ArrowLine(32,36)(64,60)
    \Photon(64,12)(64,28){-2.3}{2.5} \Vertex(64,28){1.7}\Vertex(64,12){1.7}
    \Photon(64,44)(64,60){-2.3}{2.5} \Vertex(64,60){1.7}\Vertex(64,44){1.7}
    \LongArrowArcn(64,36)(8,-20,160) \LongArrowArcn(64,36)(8,160,340)
  \end{picture}
  \qquad \text{and} \qquad \begin{picture}(40,25)(0,10)
    \Photon(0,36)(32,36){3.3}{3.5} \Vertex(32,36){1.7}
    \ArrowLine(32,36)(80,72) \ArrowLine(80,0)(32,36)
    \Photon(68,9)(48,48){-2.3}{4.5} \Vertex(68,63){1.7}\Vertex(68,9){1.7}
    \CCirc(54,36){4}{White}{White}
    \Photon(68,63)(48,24){2.3}{4.5} \Vertex(48,24){1.7}\Vertex(48,48){1.7}
  \end{picture}\,.
\end{equation}
The first one is obtained, when the sub-graph
$\SetScale{0.3}\begin{picture}(24,0)(0,7.5) \Photon(0,36)(20,36){3.3}{2.5} \Vertex(20,36){1.7}
  \Photon(60,36)(80,36){3.3}{2.5} \Vertex(60,36){1.7} \LongArrowArcn(40,36)(20,80,260)
  \LongArrowArcn(40,36)(20,260,440)
\end{picture}$ is \textit{inserted} into 
$\SetScale{0.3}\begin{picture}(24,0)(0,7.5) \Photon(0,36)(32,36){3.3}{3.5} \Vertex(32,36){1.7}
  \Line(80,72)(32,36) \ArrowLine(64,12)(32,36) \Line(80,0)(32,36) \ArrowLine(32,36)(64,60)
  \Photon(64,12)(64,60){-3.3}{4.5} \Vertex(64,60){1.7}\Vertex(64,12){1.7}
\end{picture}
$.
However, the second one does not contain any sub-graph and cannot be constructed via graph insertion. Note that
the co-product of a 1PI Feynman graph $\Gamma$ without sub-graphs is given by
$\Delta \Gamma = \Gamma \otimes \mathbb{I} + \mathbb{I} \otimes \Gamma$, see Eq.~(\ref{eq:co-product}). Hence,
$\phi_R(\Gamma) \propto L$ by the exponential formula~(\ref{eq:exp-formula}) and all graphs $\Gamma$ with that
property are called \textit{primitive graphs}. On the other hand, each non-primitive 1PI Feynman graph can be
obtained via graph insertions of primitive ones.

In the following, we make this more explicit. Therefore, let $\Gamma$ be a primitive graph. Each vertex of
$\Gamma$ represents an \textit{insertion place} for a 1PI vertex graph. Furthermore, each internal edge of type
$e$ in $\Gamma$ is an insertion place for 1PI propagator graphs with residue $e$. Note that more than one
propagator graph can be inserted into an edge, whereas only one vertex graph can be inserted into a vertex of
$\Gamma$. Now, for each primitive $\Gamma$, define an \textit{insertion operator} as a linear map
$B_+^\Gamma: \mathcal{H} \to \mathcal{H}\,$ , such that a disjoint union of 1PI Feynman graphs maps to
\begin{equation}
  \label{eq:B+G}
  B_+^\Gamma (\Gamma_1 \ldots \Gamma_n) = \sum \frac{\tilde{\Gamma}}{\text{sym}(\tilde{\Gamma})}\,.
\end{equation}
Here, the sum extends over all graphs $\tilde{\Gamma}$ that can be obtained by insertion of
$\Gamma_1,\ldots, \Gamma_n$ into the various insertion places of $\Gamma$. The factor
$\text{sym}(\tilde{\Gamma})$ is assigned such that the insertion operator commutes with the co-product in the
following way~\footnote{The $B_+^\Gamma$ are Hochschild 1-co-cycles.}:
\begin{equation}
  \label{eq:B+}
  \Delta B_+^\Gamma(\cdot) = B_+^\Gamma(\cdot) \otimes \mathbb{I} + \left(\id \otimes B_+^\Gamma\right)
  \Delta(\cdot). 
\end{equation}
The following examples should make Eq.~(\ref{eq:B+G}) clear --- for a more rigorous definition, see
\cite{Kreimer:2005rw}:
\begin{gather*}
  B_+^{\SetScale{0.15}\begin{picture}(12,0)(0,0) \Photon(0,36)(32,36){3.3}{3.5} \Vertex(32,36){1.7}
      \Line(80,72)(32,36) \ArrowLine(64,12)(32,36) \Line(80,0)(32,36) \ArrowLine(32,36)(64,60)
      \Photon(64,12)(64,60){-3.3}{4.5} \Vertex(64,60){1.7}\Vertex(64,12){1.7}
    \end{picture}} \SetScale{0.4} 
  \left(
    \begin{picture}(32,13)(0,11.5) \Photon(0,36)(20,36){3.3}{2.5} \Vertex(20,36){1.7}
      \Photon(60,36)(80,36){3.3}{2.5} \Vertex(60,36){1.7} \LongArrowArcn(40,36)(20,80,260)
      \LongArrowArcn(40,36)(20,260,440)
    \end{picture}
  \right) = \SetScale{0.5}
  \begin{picture}(40,25)(0,15.5)
    \Photon(0,36)(32,36){3.3}{3.5} \Vertex(32,36){1.7}
    \Line(80,72)(32,36) \ArrowLine(64,12)(32,36)
    \Line(80,0)(32,36) \ArrowLine(32,36)(64,60)
    \Photon(64,12)(64,28){-2.3}{2.5} \Vertex(64,28){1.7}\Vertex(64,12){1.7}
    \Photon(64,44)(64,60){-2.3}{2.5} \Vertex(64,60){1.7}\Vertex(64,44){1.7}
    \LongArrowArcn(64,36)(8,-20,160) \LongArrowArcn(64,36)(8,160,340)
  \end{picture}\,,\quad 
  B_+^{\SetScale{0.15}\begin{picture}(12,0)(0,0) \Photon(0,36)(32,36){3.3}{3.5} \Vertex(32,36){1.7}
      \Line(80,72)(32,36) \ArrowLine(64,12)(32,36) \Line(80,0)(32,36) \ArrowLine(32,36)(64,60)
      \Photon(64,12)(64,60){-3.3}{4.5} \Vertex(64,60){1.7}\Vertex(64,12){1.7}
    \end{picture}} \SetScale{0.4} 
  \left(
    \begin{picture}(32,20)(0,11.5) \Photon(0,36)(32,36){3.3}{3.5} \Vertex(32,36){1.7}
      \Line(80,72)(32,36) \ArrowLine(64,12)(32,36) \Line(80,0)(32,36) \ArrowLine(32,36)(64,60)
      \Photon(64,12)(64,60){-3.3}{4.5} \Vertex(64,60){1.7}\Vertex(64,12){1.7}
    \end{picture}\,  \right) = \SetScale{0.5}
  \frac{1}{3} \left(\begin{picture}(40,25)(0,16)
    \Photon(0,36)(32,36){3.3}{3.5} \Vertex(32,36){1.7}
    \ArrowLine(32,36)(80,72)\ArrowLine(80,0)(32,36)
    \Photon(64,12)(64,60){-3.3}{4.5} \Vertex(64,60){1.7}\Vertex(64,12){1.7}
    \Photon(48,24)(48,48){-3.3}{2.5} \Vertex(48,24){1.7}\Vertex(48,48){1.7}
  \end{picture} +
  \begin{picture}(40,25)(0,16)
    \Photon(0,36)(32,36){3.3}{3.5} \Vertex(32,36){1.7}
    \Line(80,72)(32,36) \ArrowLine(32,36)(56,54)
    \Line(80,0)(32,36) \ArrowLine(48,24)(32,36)
    \Photon(56,18)(56,54){-3.3}{4.5} \Vertex(56,54){1.7}\Vertex(56,18){1.7}
    \PhotonArc(56,18)(10,144,324){2.0}{5.5} \Vertex(48,24){1.7} \Vertex(64,12){1.7}
  \end{picture} +
  \begin{picture}(40,25)(0,16)
    \Photon(0,36)(32,36){3.3}{3.5} \Vertex(32,36){1.7}
    \Line(80,72)(32,36) \ArrowLine(32,36)(48,48)
    \Line(80,0)(32,36) \ArrowLine(56,18)(32,36)
    \Photon(56,18)(56,54){-3.3}{4.5} \Vertex(56,54){1.7}\Vertex(56,18){1.7}
    \PhotonArc(56,54)(10,36,216){2.0}{5.5} \Vertex(48,48){1.7} \Vertex(64,60){1.7}
  \end{picture} \,\right) \,.
\end{gather*}
Note that by Eq.~(\ref{eq:co-product}), these examples imply Eq.~(\ref{eq:B+}).

Some remarks are in order: First, inserting zero-loop graphs into $\Gamma$ does not change it. This is
consistent with our realization of zero-loop graphs as the unit element $\mathbb{I} \in \mathcal{H}$. For
example, $B_+^\Gamma(\mathbb{I}) = \Gamma$ and $B_+^\Gamma(\Gamma_1 \mathbb{I}) = B_+^\Gamma(\Gamma_1)$.
Secondly, inserting more than $V$ vertex graphs into a primitive with $V$ vertices does not give a well defined
object in $\mathcal{H}$. For example, there is no element in $\mathcal{H}$ that corresponds to
\begin{equation*}
    B_+^{\SetScale{0.18}\begin{picture}(12,0)(0,2)
      \ArrowLine(0,36)(80,36) \PhotonArc(40,36)(20,0,180){2.3}{7.5} \Vertex(20,36){1.7} \Vertex(60,36){1.7}
    \end{picture}} \SetScale{0.4} 
  \left(
    \begin{picture}(34,20)(0,11.5) \Photon(0,36)(32,36){3.3}{3.5} \Vertex(32,36){1.7}
      \Line(80,72)(32,36) \ArrowLine(64,12)(32,36) \Line(80,0)(32,36) \ArrowLine(32,36)(64,60)
      \Photon(64,12)(64,60){-3.3}{4.5} \Vertex(64,60){1.7}\Vertex(64,12){1.7}
    \end{picture}^3\,  \right)\,.
\end{equation*}
Finally, there are non-primitive graphs in $\mathcal{H}$ that are not in the image of some insertion operator.
For example,
$\SetScale{0.3} \begin{picture}(24,0)(0,7.5) \Photon(0,36)(32,36){3.3}{3.5} \Vertex(32,36){1.7}
  \ArrowLine(32,36)(80,72)\ArrowLine(80,0)(32,36) \Photon(64,12)(64,60){-3.3}{4.5}
  \Vertex(64,60){1.7}\Vertex(64,12){1.7} \Photon(48,24)(48,48){-3.3}{2.5} \Vertex(48,24){1.7}\Vertex(48,48){1.7}
\end{picture}$ contains a sub-graph, but it is not in the image of $
B_+^{\SetScale{0.15}\begin{picture}(12,0)(0,0) \Photon(0,36)(32,36){3.3}{3.5} \Vertex(32,36){1.7} 
    \Line(80,72)(32,36) \ArrowLine(64,12)(32,36) \Line(80,0)(32,36) \ArrowLine(32,36)(64,60)
    \Photon(64,12)(64,60){-3.3}{4.5} \Vertex(64,60){1.7}\Vertex(64,12){1.7}
  \end{picture}}$. However, the infinite sums $X^r$ in Eqs.~(\ref{eq:X-propagator}) and (\ref{eq:X-vertex})
consist of such linear combinations of 1PI Feynman graphs that are in the image of some insertion operator
$B_+^\Gamma$. This will be essential in the following.

The closed formula for the co-product in Eq.~(\ref{eq:Delta-Xr}) can be found from fixed-point relations (DSE's)
for the infinite sums $X^r$. Let us motivate this again at the QED example, where the sums of all 1PI Feynman
graphs of a certain residue satisfy \cite{Kreimer:2009iy}
\begin{equation}
\label{eq:fixed-QED-graphic}
\begin{split}
\SetScale{0.75}
\begin{picture}(60,30)(0,24.5)
\ArrowLine(5,36)(25,36) \ArrowLine(55,36)(75,36) \GCirc(40,36){15}{0.5}
\end{picture} =\, \SetScale{0.75}
\begin{picture}(60,40)(0,24.5)
\ArrowLine(0,36)(15,36)\ArrowLine(68,36)(80,36)
\ArrowLine(0,36)(80,36) \PhotonArc(40,36)(20,0,180){2.3}{7.5} \Vertex(20,36){1.7} \Vertex(60,36){1.7}
\GCirc(60,36){7}{0.5} 
\GOval(40,36)(5,8)(0){0.5} \GOval(40,54)(5,8)(0){0.5}
\end{picture}  \,+ \ldots \,, \qquad \SetScale{0.75}
\begin{picture}(60,30)(0,24.5)
\Photon(5,36)(25,36){3.3}{2.5} \Photon(55,36)(75,36){3.3}{2.5} \GCirc(40,36){15}{0.5}
\end{picture} =\, \SetScale{0.75}
\begin{picture}(60,40)(0,24.5)
\Photon(0,36)(20,36){3.3}{2.5} \Vertex(20,36){1.7} 
\Photon(60,36)(80,36){3.3}{2.5} \Vertex(60,36){1.7}
\LongArrowArcn(40,36)(20,35,215) \LongArrowArcn(40,36)(20,215,395)
\LongArrowArcn(40,36)(20,-55,125) \LongArrowArcn(40,36)(20,125,305)
\GCirc(60,36){7}{0.5}
\GOval(40,18)(5,8)(0){0.5} \GOval(40,54)(5,8)(0){0.5}
\end{picture}\, + \ldots \,,\\ \SetScale{0.75}
\begin{picture}(52,30)(0,24.5)
\Photon(0,36)(32,36){3.3}{4} \Vertex(32,36){1.7}
 \ArrowLine(42.61,46.61)(53.21,57.21)
 \ArrowLine(53.21,14.79)(42.61,25.39)
\GCirc(32,36){15}{0.5}
\end{picture} =\, \SetScale{0.75}
\begin{picture}(33,0)(0,24.5)
\Photon(0,36)(20,36){3.3}{2.5} \Vertex(20,36){1.7}
\ArrowLine(20,36)(41.21,57.21)
\ArrowLine(41.21,14.79)(20,36)
\end{picture} +
  \!\!
\begin{picture}(60,50)(0,24.5)
\Photon(10,36)(32,36){3.3}{2.5} \Vertex(32,36){1.7}
\Line(80,72)(32,36) \ArrowLine(80,0)(68,9)
\Line(80,0)(32,36) \ArrowLine(68,63)(80,72)
\Photon(64,12)(64,60){-2.3}{7.5} \Vertex(64,60){1.7}\Vertex(64,12){1.7}
\GCirc(32,36){7}{0.5}\GCirc(64,60){7}{0.5}\GCirc(64,12){7}{0.5}
\GOval(48,48)(5,8)(36.87){0.5} \GOval(48,24)(5,8)(-36.87){0.5} \GOval(64,36)(5,8)(90){0.5}
\end{picture} \,\,\,+ \ldots \hspace{80pt} \\[20pt]
\end{split}
\end{equation}
Here, we recall the notation of the introduction: a circle represents the sum of all respective 1PI graphs and
an ellipse denotes the geometric series of propagator Feynman graphs, as in
Eq.~(\ref{eq:geometric})~\footnote{Remember that the zero-loop propagator graph is contained in the ellipse, but
  not in the circle on an edge --- it belongs to the geometric series, but not to the sum of 1PI graphs.}. On
the rhs, these infinite sums are inserted into the one-loop primitives
\begin{align*}
  \SetScale{0.5}
  \begin{picture}(40,0)(0,18.5)
    \ArrowLine(0,36)(80,36) \PhotonArc(40,36)(20,0,180){2.3}{7.5} \Vertex(20,36){1.7} \Vertex(60,36){1.7}
  \end{picture} \;,\qquad
  \begin{picture}(40,0)(0,15.5)
    \Photon(0,36)(20,36){3.3}{2.5} \Vertex(20,36){1.7} 
    \Photon(60,36)(80,36){3.3}{2.5} \Vertex(60,36){1.7}
    \LongArrowArcn(40,36)(20,80,260) \LongArrowArcn(40,36)(20,260,440)
  \end{picture} \; \qquad \text{and} \qquad
  \begin{picture}(40,23)(0,15.5)
    \Photon(0,36)(32,36){3.3}{3.5} \Vertex(32,36){1.7}
    \Line(80,72)(32,36) \ArrowLine(64,12)(32,36)
    \Line(80,0)(32,36) \ArrowLine(32,36)(64,60)
    \Photon(64,12)(64,60){-3.3}{4.5} \Vertex(64,60){1.7}\Vertex(64,12){1.7}
  \end{picture} \,,\\[-7pt]
\end{align*}
where only one vertex of the propagator primitives is dressed with an infinite sum in order to avoid double
counting of graphs. The dots in Eqs.~(\ref{eq:fixed-QED-graphic}) represent all graph insertions into
higher-loop primitives. The QED example is somewhat special, because the propagator DSE can be formulated for
the \textit{quenched limit}. Then, the first two fixed-point relations in Eqs.~(\ref{eq:fixed-QED-graphic}) are
actually \textit{complete}.

Eqs.~(\ref{eq:fixed-QED-graphic}) translate to DSE's for the infinite sums $X^r$ in Eqs.~(\ref{eq:X-propagator})
and (\ref{eq:X-vertex}). These can be written in a very elegant way using the insertion operators:
\begin{equation}
\label{eq:fixed-QED}
\begin{split}
  X^{\SetScale{0.3}\begin{picture}(6,0)(0,0) \ArrowLine(0,18)(20,18)\end{picture}} = \SetScale{0.3} \mathbb{I} - B_+^{\SetScale{0.18}\begin{picture}(12,0)(0,2)
      \ArrowLine(0,36)(80,36) \PhotonArc(40,36)(20,0,180){2.3}{7.5} \Vertex(20,36){1.7} \Vertex(60,36){1.7}
    \end{picture}}
                                                                                     \left(
                                                                                     \frac{\left(X^{\begin{picture}(9,0)(0,0)
                                                                                             \Photon(0,18)(16,18){2}{2.5} \Vertex(16,18){1.7} \ArrowLine(16,18)(30,32)
                                                                                             \ArrowLine(30,4)(16,18) \end{picture}}\right)^2}{X^{\begin{picture}(6,0)(0,0)
                                                                                           \Photon(0,18)(20,18){2}{3.5}\end{picture}} X^{\begin{picture}(6,0)(0,0)
                                                                                           \ArrowLine(0,18)(20,18)\end{picture}}}
                                                                                     \right) \,,\qquad\qquad  X^{\SetScale{0.3}\begin{picture}(6,0)(0,0) \Photon(0,18)(20,18){2}{3.5}\end{picture}} = \SetScale{0.3} \mathbb{I} - B_+^{\SetScale{0.18}  \begin{picture}(12,0)(0,0)
    \Photon(0,36)(20,36){3.3}{2.5} \Vertex(20,36){1.7} 
    \Photon(60,36)(80,36){3.3}{2.5} \Vertex(60,36){1.7}
    \LongArrowArcn(40,36)(20,80,260) \LongArrowArcn(40,36)(20,260,440)
  \end{picture}} 
                                                                                          \left(
                                                                                          \frac{\left(X^{\begin{picture}(9,0)(0,0) 
                                                                                                  \Photon(0,18)(16,18){2}{2.5} \Vertex(16,18){1.7} \ArrowLine(16,18)(30,32)
                                                                                                  \ArrowLine(30,4)(16,18) \end{picture}}\right)^2}{\left(X^{\begin{picture}(6,0)(0,0)
                                                                                                  \ArrowLine(0,18)(20,18)\end{picture}}\right)^2}
                                                                                          \right) \,,\\
  X^{\SetScale{0.3}\begin{picture}(9,0)(0,0) \Photon(0,18)(16,18){2}{2.5} \Vertex(16,18){1.7}
      \ArrowLine(16,18)(30,32) \ArrowLine(30,4)(16,18) \end{picture}} = \SetScale{0.3} \mathbb{I} + B_+^{\SetScale{0.15}\begin{picture}(12,0)(0,0) \Photon(0,36)(32,36){3.3}{3.5} \Vertex(32,36){1.7}
      \Line(80,72)(32,36) \ArrowLine(64,12)(32,36) \Line(80,0)(32,36) \ArrowLine(32,36)(64,60)
      \Photon(64,12)(64,60){-3.3}{4.5} \Vertex(64,60){1.7}\Vertex(64,12){1.7}
    \end{picture}}
                                                                        \left(
                                                                        \frac{\left(X^{\begin{picture}(9,0)(0,0)
                                                                                \Photon(0,18)(16,18){2}{2.5} \Vertex(16,18){1.7} \ArrowLine(16,18)(30,32)
                                                                                \ArrowLine(30,4)(16,18) \end{picture}}\right)^3}{X^{\begin{picture}(6,0)(0,0)
                                                                              \Photon(0,18)(20,18){2}{3.5}\end{picture}} \left(X^{\begin{picture}(6,0)(0,0)
                                                                                \ArrowLine(0,18)(20,18)\end{picture}}\right)^2}
                                                                        \right) + \ldots \hspace{75pt}
\end{split}
\end{equation}
Note that there is \textit{no double counting} of Feynman graphs, although the structure differs from the one in
Eqs.~(\ref{eq:fixed-QED-graphic}). Indeed, the (negative) power of
$\SetScale{0.3} X^{\begin{picture}(9,0)(0,0) \Photon(0,18)(16,18){2}{2.5} \Vertex(16,18){1.7}
    \ArrowLine(16,18)(30,32) \ArrowLine(30,4)(16,18) \end{picture}}$
($X^e$) in the argument of an insertion operator $B_+^\Gamma$ corresponds to the number of vertices (type-$e$
edges) in $\Gamma$. Hence, in contrast to Eqs.~(\ref{eq:fixed-QED-graphic}), \textit{each} insertion place is
dressed by an infinite sum. Here, the insertion operators take care that there is no double counting and the
deeper reason for that is the commutation relation in Eq.~(\ref{eq:B+}) \cite{Kreimer:2005rw}.

With these considerations, the above relations generalize in a simple way. Therefore, denote the sum of all
insertion operators $B_+^\Gamma$ for $K$-loop primitives $\Gamma$ with residue $r$ by
\begin{align}
  \label{eq:sum-B+}
  B_+^{K;r} = \sum_{\substack{\Gamma \in \mathcal{H}_K \text{ primitive}\\ \res(\Gamma) = r}} B_+^\Gamma\,.
\end{align}
Then, the general DSE's \cite{Kreimer:2009iy,Kreimer:2006gm,Foissy_faadi,Kreimer:2006ua} for the infinite sums $X^r$ in
Eqs.~(\ref{eq:X-propagator}) and (\ref{eq:X-vertex}) with $\phi_R(r) = 1$ read
\begin{equation}
  \label{eq:DSE}
  X^{r} = \mathbb{I} + \sgn(r) \sum_{K\geq 1} B_+^{K;r} (X^{r} X_Q^K)\,.
\end{equation}
Indeed, the insertion places of each $K$-loop primitive graph are fully dressed by infinite sums~\footnote{Let
  us show this quickly: Consider a $K$-loop primitive graph $\Gamma$. The argument of the corresponding
  insertion operator $B_+^\Gamma$ is $X^r X_Q^K$, where by Eqs.~(\ref{eq:amp-greens}) and (\ref{eq:Q}),
\begin{equation*}
  X_Q = 
  \left(
    \frac{X^v}{\prod_{e \sim v} \sqrt{X^e}}
  \right)^l\,.
\end{equation*}
The power of $X^v$ in the argument of $B_+^\Gamma$ is $l\cdot K$ for $|r| = 2$ and $l \cdot K + 1$, if $r = v$.
This matches the number of vertices $V$ in $\Gamma$, given by Eq.~(\ref{eq:V-L}). Secondly, the power of
$1 \slash X^e$ in the argument of $B_+^\Gamma$ for some edge-type residue $e$ is equal to
\begin{equation*}
  l K \sum_{e \sim v} \frac{1}{2} -
  \begin{cases}
    1, & \text{if } r = e \\
    0, & \text{else}
  \end{cases} = V \sum_{e \sim v} \frac{1}{2} -
  \begin{cases}
    1, & \text{if } r = e\\ \sum_{e \sim v} \frac{1}{2}\,,& \text{if } r = v \\ 0, & \text{if } |r| = 2, \, r
    \neq e
  \end{cases}.
\end{equation*}
This equals the number of type-$e$ internal edges in $\Gamma$, which completes the proof.} and the insertion
operators ensure that there is no double counting. Note that Eqs.~(\ref{eq:fixed-QED}) coincide with this
general form, but only includes the $K=1$ terms.

To close this section, we show that the co-product formula in Eq.~(\ref{eq:Delta-Xr}) and the commutation
relation in Eq.~(\ref{eq:B+}) are consistent with the DSE in Eq.~(\ref{eq:DSE}). With slightly more effort, one
can turn this into a proof of Eq.~(\ref{eq:Delta-Xr}) by induction. For details, see e.g.
\cite{Yeats:2008zy,vanSuijlekom:2008kv,Borinsky:2014xwa,Kissler:2016gxn}. First, let
$M(X^r) = \prod_r (X^r)^{s_r}$ with integers $s_r$ be an arbitrary monomial in $X^r$. Then,
Eq.~(\ref{eq:Delta-Xr}) implies that
\begin{equation*}
  \Delta M(X^r) = \sum_{k\geq 0} M(X^r) X_Q^k \otimes M(X^r) \big|_k\,.
\end{equation*}
Now, we compute the co-product of the rhs in Eq.~(\ref{eq:DSE}). Therefore, we commute $\Delta$ with the
insertion operators via Eq.~(\ref{eq:B+}) and use the above relation for the monomials:
\begin{equation*}
  \Delta [rhs (\ref{eq:DSE})] =  X^r \otimes \mathbb{I} + \sgn(r) \sum_{K \geq 1} \sum_{k \geq 0} \left(\id
    \otimes B_+^{K;r} \right)  
  \left(
    X^r X_Q^{K+k} \otimes \left( X^r X_Q^K \right)\big|_k
  \right) \,.
\end{equation*}
Finally, we rearrange the sums as $\sum_{k\geq 1} \sum_{K = 1}^k$ using an index shift, which results in
\begin{equation*}
  \Delta [rhs (\ref{eq:DSE})] = X^r \otimes \mathbb{I} + \sum_{k \geq 1} 
  \left(
    X^r X_Q^k \otimes \sgn(r) \sum_{K = 1}^k B_+^{K;r} \left( X^r X_Q^K \big|_{k-K} \right)
  \right) \,.
\end{equation*}
Note that the terms on the rhs of the $\otimes$ sign correspond to $X^r |_k$ (see Eq.~(\ref{eq:DSE})). Hence, we
are left with
\begin{equation*}
  \Delta [rhs (\ref{eq:DSE})] = X^r \otimes \mathbb{I} + \sum_{k \geq 1} 
  X^r X_Q^k \otimes X^r \big|_k = \sum_{k\geq 0} X^r X_Q^k \otimes X^r \big|_k = \Delta [lhs (\ref{eq:DSE})]\,.
\end{equation*}

\subsection{Rederivation of the Callan-Symanzik Equation}
\label{sec:derivation-1}

With the considerations of the previous sections, the proof of Eq.~(\ref{eq:MAIN}) is simple. It requires the
exponential formula~(\ref{eq:exp-formula}) and the co-product relation~(\ref{eq:Delta-Xr}). Together with the
definitions of the convolution product in Eq.~(\ref{eq:star}) and the $\sigma$ function in
Eq.~(\ref{eq:sigma-anom}), we find
\begin{align*}
  \frac{\partial G^r}{\partial L} = \frac{\partial}{\partial L} \, \phi_R(X^r) = \frac{\partial}{\partial L} \,
  \exp^\star(L \sigma) (X^r) &= (\phi_R \star \sigma) (X^r) = \\[5pt] = \sum_{k\geq0} \phi_R \left(X^r X^k_Q \right)
  \sigma \left(X\big|_k \right) &= \sum_{k\geq 1} G^r Q^k \left( - \gamma_{k - 1}^r \alpha^k \right) = - G^r
  \gamma^r(\alpha Q)\,.
\end{align*}

\section{Multiple Coupling Constants}
\label{sec:gener-results}

In this section, we rederive the Callan-Symanzik equation from Dyson-Schwinger equations in QFT's with multiple
interaction terms in the Lagrangian. Therefore, we assume that each such term comes with a separate coupling
constant and gives rise to a certain vertex in Feynman graphs. We denote the different vertices by
$v_1,\ldots,v_m$ and collect the coupling parameters to a vector $g = (g_1\ldots g_m)$. The positive integer
$m$ is given by the number of interaction terms. In the following, we summarize the main changes to the
one-coupling case.

First of all, it is inappropriate to grade the Hopf algebra $\mathcal{H}$ of Feynman graphs with respect to the
loop number in this setting. Instead, $\mathcal{H}$ is multi-graded as
\cite{Connes:1999yr,Foissy:2015wqu,Prinz:2018dhq,Prinz}
\begin{equation*}
  \mathcal{H} = \bigoplus_\mathbf{k} \mathcal{H}_\mathbf{k}\,.
\end{equation*}
Here, $\mathbf{k} = (k_1,\ldots,k_m)$ denotes a multi-index (a vector of integers) and the sub-spaces
$\mathcal{H}_\mathbf{k}$ contain the following Feynman graphs:
\begin{equation*}
  \Gamma \in \mathcal{H}_\mathbf{k} \qquad \Leftrightarrow \qquad
  \Gamma \text{ has }\begin{cases}
     \quad \, k_j  \quad\, \text{ vertices } v_j \,,& \text{if } |\res\,\Gamma| = 2 \\
     k_j + \delta_{ij} \text{ vertices } v_j \,,& \text{if }\,\, \res\,\Gamma = v_i 
  \end{cases}\,.
\end{equation*}
Note that this grading respects the product and co-product of $\mathcal{H}$.

Secondly, the Green's functions $G^r$ depend on all coupling constants $g$ and the logarithmic energy scale $L$.
In particular, $G^r(g,L) = \phi_R(X^r)$ as before, where the $X^r$ are given by the infinite sums in
Eqs.~(\ref{eq:X-propagator}) and (\ref{eq:X-vertex}). Hence, the corresponding anomalous dimensions $\gamma^r$
are power series in all coupling parameters,
\begin{equation*}
  \gamma^r(g) = - \frac{\partial G^r(g,L)}{\partial L} \bigg|_{L = 0} = \sum_\mathbf{k}
  \gamma^r_\mathbf{k} \, g_1^{k_1} \ldots g_m^{k_m}\,.
\end{equation*}

Furthermore, there is one amputated Green's function (invariant charge) for each vertex $v_i$, see
Eq.~(\ref{eq:amp-greens}). These define the running coupling parameters as before. In order to simplify the
notation, we write
\begin{equation*}
  X_i := X_\amp^{v_i}\,, \qquad Q_i(g,L) := \phi_R(X_i)\,,\qquad \tilde{g} = (\tilde{g}_1\ldots \tilde{g}_m) =
  (g_1 Q_1 \ldots g_m Q_m)\,. 
\end{equation*}
Accordingly, there is one $\beta$-function for each vertex $v_i$, with power series expansion
\begin{equation*}
  \beta^i(g) = \frac{\partial Q_i(g,L)}{\partial L} \bigg|_{L = 0} = \sum_\mathbf{k}
    \beta^i_\mathbf{k} \,g_1^{k_1} \ldots g_m^{k_m}\,.
\end{equation*}

Finally, the Dyson-Schwinger equation for the infinite sums in Eqs.~(\ref{eq:X-propagator}) and
(\ref{eq:X-vertex}) is given by
\begin{equation}
  \label{eq:DSE-multi}
  X^r = \mathbb{I} + \sgn(r) \sum_\mathbf{k} B_+^{\mathbf{k} ; r}
  \left(X^r X_1^{k_1} \ldots X_m^{k_m}\right)\,,
\end{equation}
where
\begin{equation*}
  B_+^{\mathbf{k} ; r} = \sum_{\substack{\Gamma \in \mathcal{H}_\mathbf{k} \text{ primitive}\\ \res(\Gamma) =
      r}} B_+^\Gamma 
\end{equation*}
is the generalization of Eq.~(\ref{eq:sum-B+})~\footnote{This relation is barely written down properly in the
  literature, see e.g. \cite{Foissy:2015wqu,Lucia}}. Note that in Eq.~(\ref{eq:DSE-multi}), each insertion place
of a primitive $\Gamma$ is fully dressed with an infinite sum and there is no double-counting of graphs because
the insertion operators $B_+^{\mathbf{k};r}$ are Hochschild 1-co-cycles, i.e. they satisfy the commutation
relation in Eq.~(\ref{eq:B+}). In full analogy to the end of Section~\ref{sec:graph-insertion}, the DSE in
Eq.~(\ref{eq:DSE-multi}) implies the co-product formula
\begin{equation}
  \label{eq:delta-Xr-multi}
  \Delta X^r = \sum_\mathbf{k} X^r X_1^{k_1} \ldots X_m^{k_m} \otimes X^r \big|_{\mathbf{k}}\,,
\end{equation}
where $\big|_\mathbf{k}$ is the projection onto $\mathcal{H}_\mathbf{k}$. This is the generalization of
Eq.~(\ref{eq:Delta-Xr})\,. 

The Callan-Symanzik equation can now be derived in the same way as in Section~\ref{sec:derivation-1}. Therefore,
the co-product formula in Eq.~(\ref{eq:Delta-Xr}) is replaced by Eq.~(\ref{eq:delta-Xr-multi}) and the action of
$\sigma$ on the infinite sums $X^r$ in Eq.~(\ref{eq:sigma-anom}) must be generalized to
\begin{equation*}
  \sigma \left( X^r \big|_\mathbf{k} \right) = - \gamma^r_\mathbf{k} \, g_1^{k_1} \ldots g_m^{k_m}\,.
\end{equation*}
Then, using the exponential formula in Eq.~(\ref{eq:exp-formula}) and the definition of the convolution product
in Eq.~(\ref{eq:star}) results in
\begin{equation*}
  \frac{\partial G^r}{\partial L} = (\phi_R \star \sigma) (X^r) =
  G^r \sum_\mathbf{k} Q_1^{k_1} \ldots Q_m^{k_m} \sigma \left( X^r\big|_\mathbf{k} \right) = - G^r
  \gamma^r(\tilde{g})\,,
\end{equation*}
which can be written in the same form as in the one-coupling case, see Eq.~(\ref{eq:MAIN}):
\begin{equation}
  \label{eq:MAIN-multi}
  \frac{\partial}{\partial L} \log G^r(g,L) = - \gamma^r(\tilde{g})\,, \qquad \tilde{g} = (g_1 Q_1 \ldots g_m
  Q_m)\,. 
\end{equation}
In particular, taking certain linear combinations of these relations yields a system of ordinary differential
equations, which is the analogue to Eq.~(\ref{eq:dQdL}),
\begin{equation}
  \label{eq:dQdL-multi}
  \frac{\partial}{\partial L} \log Q_i(g,L) = \beta^i(\tilde{g})\,, \qquad \tilde{g} = (g_1 Q_1 \ldots g_m
  Q_m)\,.
\end{equation}
Together with the initial conditions $G^r(g,0) = 1$ and $Q_i(g,0) = 1$, Eqs.~(\ref{eq:MAIN-multi}) and
(\ref{eq:dQdL-multi}) fully determine the Green's functions, when only the anomalous dimensions and
$\beta$-functions are known.

Two remarks are in order. First, in terms of the running coupling parameters, Eq.~(\ref{eq:dQdL-multi})
represents the usual system of renormalization group equations~\footnote{Again, our definition of the
  $\beta$-functions differs from the one in the literature,
  $\beta^i_\text{lit}(\tilde{g}) = \tilde{g}_i \beta^i(\tilde{g})$.} and Eq.~(\ref{eq:MAIN-multi}) is the usual
form of the Callan-Symanzik equation for the Green's functions $\tilde{G}^r$ with
$\tilde{G}^r(\tilde{g},L) = G^r(g,L)$:
\begin{equation*}
  \frac{\partial \tilde{g}_i}{\partial L} = \tilde{g}_i \beta^i(\tilde{g})\,,\qquad
  \left(
    \frac{\partial}{\partial L} + \sum_{i} \tilde{g}_i \beta^i(\tilde{g}) \frac{\partial}{\partial
      \tilde{g}_i} + \gamma^r(\tilde{g})
  \right) \tilde{G}^r(\tilde{g},L) = 0\,.
\end{equation*}

Secondly, the log-expansions for the Green's functions can be derived as in Section~\ref{sec:log-expansion} via
a change of variables $L \to z = g_i L$ (for one of the coupling parameters $g_i$). In particular, let
$G^r(g,L) = H^r(g, g_i L)$. Then, the solution to Eq.~(\ref{eq:MAIN-multi}) is given by
\begin{equation}
  \label{eq:H-exp-multi}
  H^r(g,z) = \exp 
  \left(
    - \int_0^z \frac{\gamma^r(\tilde{g})}{g_i} \,\mathrm{d}z'
  \right)\,.
\end{equation}
This formula only requires the log-expansion for the running coupling parameters $\tilde{g}$, which can be found
log order by log order from the RGE. We hope to investigate this case further and give examples in future work.

\section{Conclusions}
\label{sec:conclusions}

In this paper, we established a precise connection between Dyson-Schwinger equations for the Green's functions
in a given QFT and their corresponding log-expansions. We mainly discussed QFT's with only one interaction term
in the Lagrangian (Section~\ref{sec:log-expansion}), but also expanded the results to the more general case
(Section~\ref{sec:gener-results}).

The formulas for the log-expansions are given in Eqs.~(\ref{eq:exp-greens-function}) and (\ref{eq:H-exp-multi})
--- the next-to$^{(n)}$-leading log approximation is obtained by an analytical integration that requires the
solution of the renormalization group equation up to the next-to$^{(n)}$-leading log order. In particular, it
only depends on the $(n+1)$-loop $\beta$-function and anomalous dimension. An explicit expression for the
next-to-next-to-leading log approximation is given in Eq.~(\ref{eq:NNLL-H}), when only one coupling constant is
involved. In particular, for the photon propagator Green's function in quantum electrodynamics (and in a toy
model, where all Feynman graphs with vertex sub-divergences are neglected) our formulas lead to the known
expressions in the literature (Compare Eq.~(\ref{eq:H_s}) with \cite{Yeats}).

Our results may be used for an accurate perturbative calculation of the Green's functions, for energy scales far
away from the renormalization scale $S \gg S_0$. However, the log-expansions do only converge, if the
coefficients do well behave perturbatively. The latter must not be true --- so far, we have no control about the
dependence of the Green's functions on the scattering angles $\mathbf{\Theta}$ and $\mathbf{\Theta}_0$. In order
to understand this better, one needs to generalize the exponential formula in Eq.~(\ref{eq:exp-formula}) to
include also the angle dependence. Together with the co-product formulas in Eqs.~(\ref{eq:Delta-Xr}) and
(\ref{eq:delta-Xr-multi}), this would lead to a generalization of the Callan-Symanzik equation, from which the
correct perturbative approach may be obtained. However, this is a highly non-trivial problem, as it requires to
understand the behavior of the Green's functions orthogonal to the renormalization group flow .

Our work gives the derivation of the log-expansions for the Green's functions from the corresponding
Dyson-Schwinger equations. However, if the latter are not known, we cannot comment on the corresponding
log-expansions. For example in QED, a detailed Dyson-Schwinger equation for the 4-point function
$\SetScale{0.3} G^{\begin{picture}(9,0)(0,0) \ArrowLine(16,18)(2,32) \ArrowLine(2,4)(16,18) \Vertex(16,18){1.7}
    \ArrowLine(16,18)(30,32) \ArrowLine(30,4)(16,18) \end{picture}}$
in terms of the insertion operators is not explicitly written down in the literature. We plan to do that in
future work --- with our analysis, this would lead to the respective Callan-Symanzik equation and then, to the
corresponding log-expansion.

Finally, we would like to make the case with multiple coupling parameters more explicit. Therefore, we need to
solve the system of renormalization group equations in Eq.~(\ref{eq:dQdL-multi}) analytically, such that
Eq.~(\ref{eq:H-exp-multi}) can be integrated log-order by log-order.

\section*{Acknowledgments}

First of all, I want to thank Dirk Kreimer and Karen Yeats for very illuminating discussions and for their
patience in answering so many questions. Furthermore, I would like to thank Simon Pl\"atzer and Angelika Widl
for giving me insights to the particle physicist point of view. Finally, I acknowledge useful discussions with
Henry Kissler, David Prinz, Stefan Fredenhagen, Michael Borinsky and John Gracey.

\bibliography{literature}{}
\bibliographystyle{JHEP}

\end{document}